\documentclass[%
 reprint,
superscriptaddress,
aip,
]{revtex4-1}
\usepackage{soul}
\usepackage{graphicx}
\usepackage{dcolumn}
\usepackage{bm}
\usepackage{hyperref}
\usepackage[mathlines]{lineno}
\usepackage{xcolor}
\usepackage{enumitem}
\usepackage{amsmath}
\usepackage{microtype}
\begin{document}

\title{Atomistic-Level Analysis of Nanoindentation-Induced Plasticity in 
Arc--Melted NiFeCrCo Alloys: The role of stacking faults}

\author{F. J. Dom\'inguez-Guti\'errez}
\author{A. Olejarz}
\affiliation{%
NOMATEN Centre of Excellence, National Center for Nuclear Research, 
05-400 Swierk/Otwock, Poland
 }%
 
\author{M. Landeiro Dos Reis}
\affiliation{LaSIE UMR CNRS 7356, La Rochelle Université, 
Av. Michel Crépeau, 17042, La Rochelle Cedex 1, France}
\author{E. Wyszkowska}
\author{D. Kalita}
\author{W. Y. Huo}
\author{I. Jozwik}
\author{L. Kurpaska}
\author{S. Papanikolaou}
\affiliation{%
NOMATEN Centre of Excellence, National Center for Nuclear Research, 
05-400 Swierk/Otwock, Poland
 }%
\author{M. J. Alava}
\affiliation{%
NOMATEN Centre of Excellence, National Center for Nuclear Research, 
05-400 Swierk/Otwock, Poland
 }%

 \author{K. Muszka}
 \affiliation{
 AGH University of Science and Technology, al. Mickiewicza 30, 30--059 Kraków, Poland
 }
\date{\today}
\begin{abstract}

Concentrated solid solution alloys (CSAs) have attracted 
attention for their promising properties; however, current manufacturing 
methods face challenges in complexity, high costs, and limited scalability, 
raising concerns about industrial viability. The prevalent technique, arc 
melting, yields high-purity samples with complex shapes. In this study, we 
explore nanoindentation tests at room temperature where arc-
melted samples exhibit larger grain sizes, diminishing the effects of grain boundaries on the results.
Motivated by these findings, our investigation focuses on the atomistic-level exploration of plasticity mechanisms, specifically dislocation 
nucleation and propagation during nanoindentation tests. The intricate 
chemistry of NiFeCrCo CSA influences pile-ups and slip traces, aiming to 
elucidate plastic deformation by considering both pristine and pre-existing 
stacking fault tetrahedra. Our analysis scrutinizes dynamic deformation processes, defect nucleation, and evolution, complemented by stress-strain and dislocation densities-strain curves illustrating the hardening mechanism of defective materials.
Additionally, we examine surface morphology and plastic deformation through 
atomic shear strain and displacement mappings. This integrated approach provides insights into the complex interplay between material structure and mechanical behavior, paving the way for an enhanced understanding and potential advancements in CSA applications.
\end{abstract}

\maketitle
\section{\label{sec:intro}Introduction}

High- and medium-entropy alloys (HEAs/MEAs) are a focal
point of research due to the extensive variety of available
elements, offering adaptable deformation behaviors 
\cite{WAGNER2022117693,li2017interstitial,HUO2017125,SHAHMIR2017342,cichocki2022effect,ma14195764}. 
While HEAs/MEAs have sparked substantial interest, systematic
studies on their mechanical properties, particularly in 
well--controlled microstructures, have been nowadays an
interesting topic for applications at extreme operating
environments 
\cite{GEORGE2020435,HeNanoindentation,MusicalAPL,YehHEA,BABILAS2023170839}. 
Despite challenging comparisons with conventional alloys, HEAs exhibit 
similar deformation mechanisms, offering a foundation to develop advanced 
alloys \cite{MIRACLE2017448,WenyiAPL,OLEJARZ2023168196,KARIMI2023115559}. 
HEAs produced through arc melting technique after some plastic and heat treatments showcase highly promising characteristics across diverse applications, often exceeding initial expectations.
The lattice distortion strengthening
effect, typical for HEAs, further contributes to 
interstitial--vacancy recombination, reinforcing the alloy's
radiation resistance \cite{OLEJARZ2023168196}.
By bridging knowledge gaps, 
the community can unlock the potential of HEAs, achieving 
a harmonious synergy of multiple mechanisms for improved
mechanical properties essential for structural applications 
\cite{KURPASKA2022110639,FRYDRYCH2023104644,GEORGE2020435,ma14195764,stasiak2022effects}.

A category of materials referred to as single--phase concentrated 
solid--solution alloys (SP--CSAs), including High Entropy Alloys (HEAs)
\cite{lu2016enhancing,JIN201665,KURPASKA2022110639}, 
has recently attracted considerable attention due to their
distinctive structures and exceptional properties. 
Unlike traditional alloys, SP--CSAs are composed of two to five
principal elements in nearly equal molar ratios, forming random solid 
solutions in either a simple face-centered cubic crystal lattice 
structure \cite{DARAMOLA2022111165}. 
The intricate random arrangement of alloying elements and the
local chemical environment at the atomic level contribute to
the extraordinary properties exhibited by SP--CSAs compared
to conventional alloys \cite{lu2016enhancing,JIN201665}.
An exemplary four-element FeCrCoNi alloy \cite{ZHONG2023103663}
is recognized for its superior strength compared to FeNiCoMn. 
This alloy demonstrates impressive tensile strength and ductility 
achieved through sustained work hardening, and its fabrication holds 
significance for applications in extreme operating environments 
\cite{OLEJARZ2023168196} being the material of interest in our work. 
The exceptional characteristics of SP-CSAs, as exemplified by
this specific alloy, underscore their potential significance
in advancing materials science for a range of demanding
applications \cite{ZHONG2023103663,OLEJARZ2023168196}.

Nanoindentation serves as a crucial technique for assessing the 
mechanical properties of potential candidate materials at
the nanoscale, particularly for applications in extreme
operating conditions. This method involves inserting a small, 
sharp tip into the sample surface to measure the force required
and the resulting displacement within the material 
\cite{SCHUH200632,VARILLAS2017431,KURPASKA2022110639,PATHAK20151,YANG2022143685,PATHAK2016241,FRYDRYCH2023104644}. 
The occurrence of a pop-in event, marking the transition from
elastic to plastic deformation as the indenter tip penetrates
the surface, is often used as a reference point for analyzing
a material's mechanical behavior \cite{PATHAK20151}. 
This event provides insights into the internal structure of
the sample, allowing for an exploration of the mechanisms
responsible for plastic deformation initiation and subsequent
modification of the material's mechanical properties 
\cite{REMINGTON2014378,D2NR06178C}.
Furthermore, nanoindentation-induced plastic patterning is
a process that creates patterns or structures at the nanoscale,
with applications ranging from fabricating nanostructured
surfaces with enhanced functionalities to developing new
materials with tailored mechanical properties \cite{Li_2020}.
The resulting plastic patterning is influenced by factors such
as applied load, crystal orientation, temperature,
and material properties, requiring a fundamental understanding
of material deformation within the plastic zone beneath
the indented surface region 
\cite{VARILLAS2017431,DOMINGUEZGUTIERREZ202238}. 
However, the mechanical response of materials is inherently tied to the 
extensive array of defects present within the material. Models can aid in 
decoupling the effects observed experimentally, although they present a challenge 
since modeling sufficiently large systems is complex to account for a diverse 
distribution of defects. Among these defects, stacking fault tetrahedra form more 
easily in High Entropy Alloys (HEA) compared to conventional alloys due to the 
lower stacking fault energy in these alloys 
\cite{hu2022formation,chandan2021temperature,huang2015temperature}. Hence, in 
this paper, we choose to focus on this peculiar defect. We investigate both 
scenarios, with and without stacking fault tetrahedra, to examine the impact of 
this defect on the mechanical response of HEA.

In our comprehensive study, we conducted both experimental and 
computational investigations on NiFeCrCo high entropy alloys
being prepared using the arc melting technique, followed by high-
temperature homogenization process, which is the common methodology 
for high-entropy alloys' manufacturing.
In the computational aspect of our study, we considered the composition 
derived from experimental data along with preexisting stacking faults 
tetrahedra. This allowed us to model the nanomechanical response of 
defected samples during nanoindentation tests.
Our computational analysis involved reporting dislocation densities, 
mechanical responses through stress-strain curves, and examining the 
surface morphology of indented samples \cite{DOMINGUEZGUTIERREZ202238}. 
This integrated approach provides a comprehensive understanding of the 
material's behavior under nanoindentation conditions.

\section{Experimental methods}

In our experimental investigation, a composition of Co, Cr, Fe, and Ni 
pieces were mixed in Ar-glovebox, followed by melting using an Edmund 
Buehler AM200 arc melting device equipped with Copper-mould. The furnace 
was evacuated up to 5$\times10^{-5}$ mbar and backfilled with Argon gas to 
a pressure of 600 mbar. To ensure the homogeneity, sample was flipped and 
remelted at least six times. Sample was homogenized in muffle furnace at 
1200°C for 4 hours to remove dendritic structure and to promote uniform 
distribution of all elements present. We characterize the obtained 
NiFeCrCo sample via XRF technique obtaining a chemical composition (in 
wt.\%) of 26 \% Ni, 25 \% Fe, 26 \% Co, and 22 \% Cr. The amount of all 
impurities did not exceed 1 \%.To characterize mechanical properties, 
nanoindentation using a NanoTest Vantage system with a Synton-MDP diamond 
Berkovich-shaped indenter was performed. Tests used a 150 mN load, with 100 
indentations on sample at 100 µm spacing.
\subsection{Structural characterization}

The microstructural evolution of homogenized sample was investigated using 
the ThermoFisher Helios 5 UX scanning electron microscope. This advanced 
microscope is equipped with the EDAX Velocity Pro EBSD camera, enabling 
precise crystal orienation mapping. EBSD mapping was performed with a 0.25 
$\mu$m step size. The acquired data underwent thorough analysis using the 
EDAX OIM Analysis 8 software, with points having a confidence index (CI) 
below 0.1 being excluded from calculations.

\section{Computational methods}
\label{sec:MLIP}

To perform our simulations, we use the Large-scale Atomic/Molecular
Massively Parallel Simulator (LAMMPS) software \cite{THOMPSON2022108171},
which allows us to study the behavior of materials under a wide range 
of conditions. 
One of our goals is to accurately model plastic deformation, which is a 
crucial aspect of how materials respond to external loads. 
Thus, we utilize interatomic potentials reported by 
Choi et al. \cite{Choietal}; which 
are based on the second Nearest Neighbor Modified Embedded Atom
Method (2NN-MEAM) as reported in our previous work \cite{KURPASKA2022110639,FRYDRYCH2023104644,naghdi2022dynamic}.




We initially defined the FCC Ni sample with crystal orientations
along [001], [101], and [111], followed by the optimization of 
the system's energy using The FIRE (Fast Inertial Relaxation 
Engine) 2.0 protocol \cite{GUENOLE2020109584} to identify 
the lowest energy structure
\cite{KURPASKA2022110639,DOMINGUEZGUTIERREZ202238,DOMINGUEZGUTIERREZ2021141912,XU2024112733}. 
The equilibration continued until the system achieved homogeneous
temperature and pressure profiles at a density of 8.23 $g/cm^3$, 
in good agreement with the experimental value \cite{D2NR06178C}. 
A final step involved relaxing the prepared sample for 10 ps 
to dissipate any artificial heat.
To create the FCC CSA, randomness was introduced by 
substituting 75\% of Ni atoms from the original sample,
oriented in different crystal directions, for 26 \% Ni, 25 \% Fe,
26 \% Co, and 22 \% Cr in wt, following the experimental results.
The optimization process for the samples' geometry is directed
towards the nearest
local minimum of the energy structure. The optimization
criteria include ensuring that the change in energy
between successive iterations and the most recent energy
magnitude remains below 10$^{-5}$. 
Additionally, the global force vector length of all
atoms is maintained at less than or equal
to $10^{-8}$ eV/\AA{}.
The subsequent steps for preparing the sample mirrored
those of the pristine case, resulting in an equiatomic
NiFeCrCoMn CSA.

\begin{figure}[t!]
    \centering
    \includegraphics[width=0.48\textwidth]{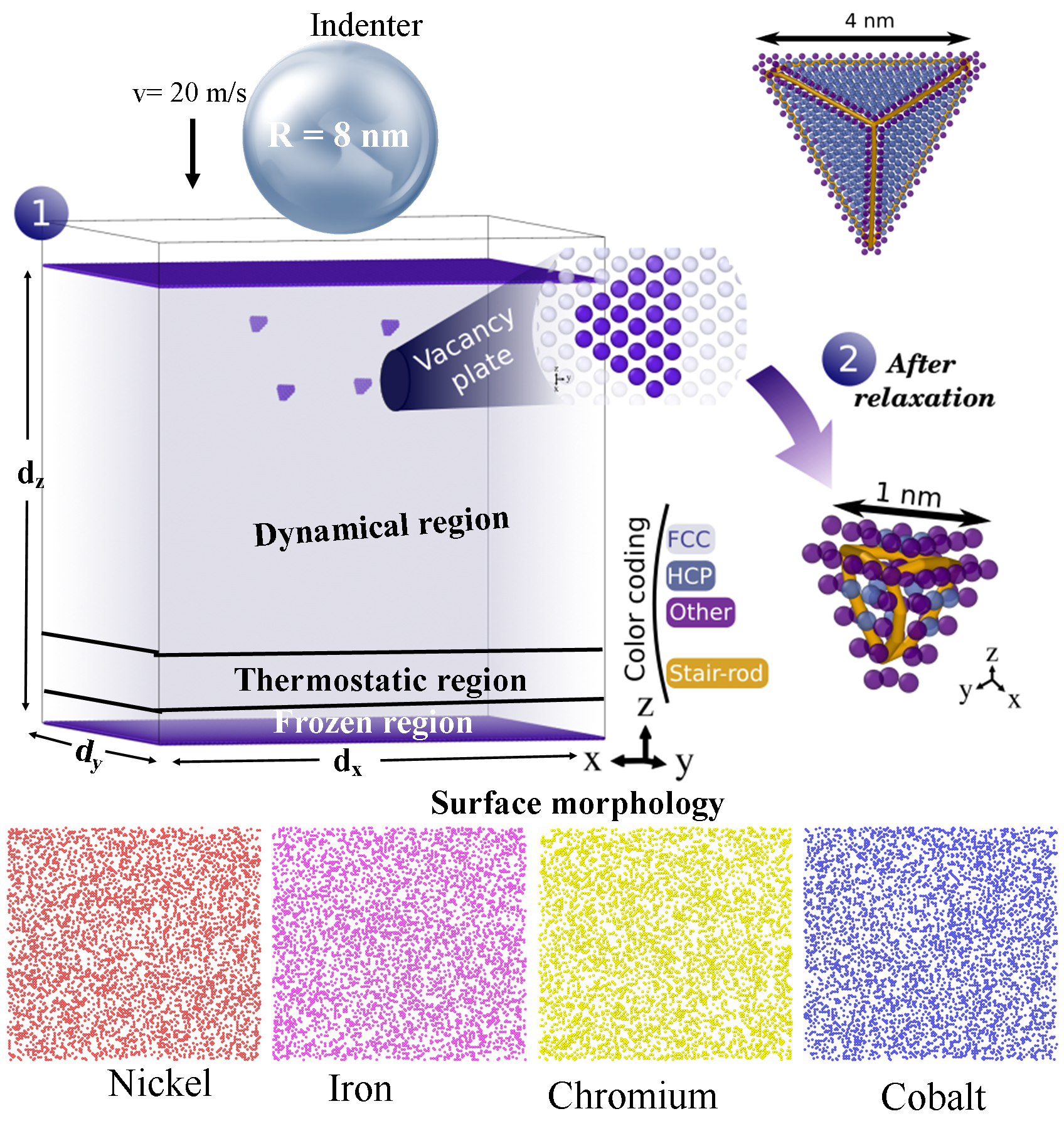}
    \caption{ (Color online) Simulation cell exhibiting four vacancy plates, each measuring 1 nm in size. Upon relaxation, these plates undergo transformation into a stacking fault tetrahedra (SFT). Additionally, a 4 nm-sized SFT is included for comparison. Initial frame of the nanoindentation simulation, 
    emphasizing the stacking fault tetrahedron and stair--rod 
    dislocations near the surface. To enhance visual clarity, 
    an atom with an FCC lattice position has been removed. 
    The surface top view showcases the atomic arrangement of 
    the five types of atoms: Ni, Fe, Cr, Co, and Mn and the indents. 
    Additionally, the sample is strategically divided into three 
    sections--frozen, thermostatic, and dynamical 
    to facilitate the mechanical test.}
    \label{fig:Nanoinitial}
\end{figure}

In agreement with the Silcox and Hirsch mechanism, stacking fault 
tetrahedra (SFT) can be formed from an equilateral triangular vacancy 
plate situated on the (111) planes \cite{silcox1959direct}.
To construct HEA-CSA with pre-existing SFT, we initiate the 
process by introducing four vacancy plates within the (111)
planes beneath the surface (See Fig. \ref{fig:Nanoinitial}.) 
\cite{PhysRevMaterials.4.103603}.
The subsequent energy minimization prompts the plates to undergo 
transformation into SFTs, characterized by stair--rod dislocations 
(depicted by the yellow line on the SFTs Fig. \ref{fig:Nanoinitial}.)
and a hexagonal close-packed (HCP) structure between
these dislocations (atoms colored in blue Fig. \ref{fig:Nanoinitial}.).
Various scenarios were examined to investigate the impact of
SFT size and distribution. The configurations considered
include one without any SFT, followed by configurations
with four SFTs of 1 nm size, four SFTs of 4 nm size, and,
lastly, four SFTs with sizes of 1 nm, 2 nm, 3 nm, and 4 nm,
respectively.


\begin{table}[b!]
\centering
\caption{Size of the initial
numerical samples used to perform MD simulations of pristine 
HESSA \cite{FRYDRYCH2023104644}. 
Sample size ($d_{x}$, $d_{y}$, $d_{z}$) in units of nm.}
\label{tab:MD_data}
\begin{tabular}{crrr}
\hline
Orientation & [001] & [011] & [111] \\
\hline
$d_{x}$ & 33.20 & 33.06 & 33.07 \\
$d_{y}$ & 36.09 & 36.17 & 36.12 \\
$d_{z}$ & 36.12 & 38.17 & 40.61 \\
Atoms & 3 164 800 & 3 442 032 & 3 645 720 \\
X-axis & $(100)$ & $(100)$ & $(\overline{1}01)$ \\
Y-axis & $(010)$ & $(01\overline{1})$ & $(1\overline{2}1)$ \\
Z-axis & $(001)$ & (011) & $(111)$ \\
\hline
\end{tabular}
\end{table}

During the initial stage, the CSA samples underwent division
into three distinct sections along the $z$ direction to establish 
boundary conditions 
along their depth, denoted as $d_z$: 1. Frozen Section: A stabilizing
frozen section, approximately 0.02$\times d_z$ in width, was
implemented to ensure the numerical cell's stability. 2.Thermostatic
Section: Positioned roughly 0.08$\times d_z$ above the frozen section,
this thermostatic section was dedicated to dissipating heat
generated during the nanoindentation process. 3. Dynamical Atoms
Section: This section allowed for dynamic interaction with the indenter
tip, resulting in modifications to the surface structure of the samples.
Additionally, a 5 nm vacuum section was incorporated at the top of
the sample, serving as an open boundary 
\cite{KURPASKA2022110639,DOMINGUEZGUTIERREZ2021141912,PhysRevMaterials.7.043603}.

Within the spatial (few nanometers) and temporal (picoseconds)
scales of the MD simulations, nanoindentation testing allows for the 
examination of the initial stages of plastic deformation and phase 
transformation at the atomic level. In the numerical modeling, 
a Berkovich tip can be effectively represented by a spherical
indenter tip due to the minimal distinctions observed in the
phase transformation regions of both tips during the early
stages of the loading process \cite{VARILLAS2017431, KURPASKA2022110639}.
The indenter tip is defined as a non--atomic repulsive
imaginary (RI) rigid sphere, exerting a force of magnitude
$F(t) = K \left( r(t) - R_i \right)^2$ on each atom. 
In this equation, $K = 200$ eV/\AA{}$^3$ denotes the 
specified force constant, ensuring high stiffness for our 
indenter tip \cite{DOMINGUEZGUTIERREZ2021141912}. 
The parameter $r(t)$ represents the distance from the 
atoms to the center of the indenter, and $R_i = 8$ nm 
is the radius of the indenter, chosen to be sufficiently 
large to accurately model the elastic--to--plastic
deformation transition \cite{PhysRevMaterials.7.043603}.
The initial location of the tip is set at a separation distance 
of 0.5 nm from the material's surface, and its center is 
randomly shifted to ten different positions to account for 
statistical variation, as depicted by black dots in Fig. 
\ref{fig:Nanoinitial}.
We employ an NVE statistical thermodynamic ensemble with the 
velocity Verlet algorithm for an indenter speed of $v = 20$ m/s 
for 125 ps, using a time step of $\Delta t = 1$ fs. A maximum
indentation depth of $h_{\rm max}=2.0$ nm is selected to minimize
boundary layer effects in the dynamic atoms region 
\cite{KURPASKA2022110639}.
The initial configuration of the nanoindentation simulation is
illustrated in Fig. \ref{fig:Nanoinitial}. 





\section{\label{sec:results}Results}

\begin{figure}[b!]
    \centering
    \includegraphics[width=0.48\textwidth]{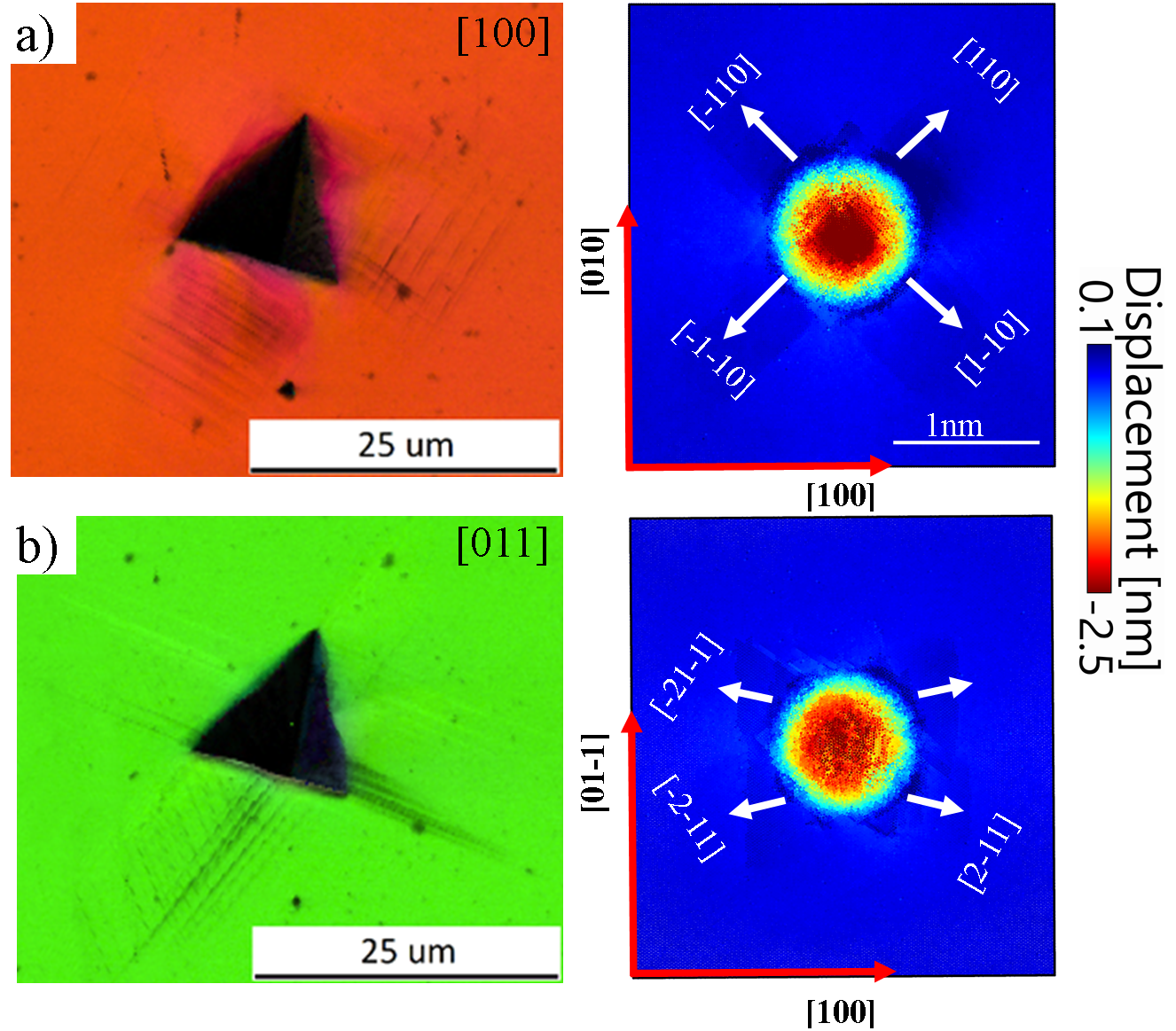}
    \caption{(Color online) EBSD image of (a) [001] and (b) [011] orientation identified by the IPF mapping.
    The images revealing crystallographic effects on the plastic 
    deformation of the polycrystalline NiFeCrCo alloy. The image showcases 
    the propagation of pile--ups across different slip planes, providing insights obtained from mapping 100 nanoindents in the experimental nanoindentation tests.
    Observation of slip--traces
    and pile--ups formation at a maximum indentation depth of
    3 nm for the [100] and [011] orientations obtained by MD simulations 
    are presented next to experimental images. The propagation of slip traces is indicated by
    their corresponding planes, exhibiting typical rosette shapes 
    characteristic of FCC metals.
    }
    \label{fig:ExpResult}
\end{figure}

 We conducted experimental nanoindentation tests on a polycrystalline 
 NiFeCrCo alloy, with a detailed examination involving the mapping
 of 100 nanoindents. 
 The investigation extends beyond the mechanical tests to explore 
 crystallographic effects on the plastic deformation of the material. 
 Utilizing EBSD imaging, we obtained 
 insights into the propagation of pile--ups 
 within various slip planes. The EBSD images of the [001]
 and [101] orientations are presented in Fig 
 \ref{fig:ExpResult}a-b). 
 The emission of surface slip traces is a crucial manifestation of crystal 
 plasticity, providing essential insights for a comprehensive understanding 
 of the mechanical response of materials. EBSD analysis of the samples 
 reveals slip traces for $\{110\}$ indenter planes in $\langle112\rangle$ directions 
 (Fig. \ref{fig:ExpResult}(b)).
 A close-up of the [101] 
 orientation of the alloy provides additional detail, showcasing the 
 propagation of slip traces on \{211\} planes—features typically
 observed in FCC (Face-Centered Cubic) materials 
 and in a good agreement with our MD simulations. 
 This arises from the interaction between dislocation gliding in $\{111\}$ 
 planes and the indenter surfaces in the $\{110\}$ plane 
 \cite{VARILLAS2017431}. Similarly, when considering $\{100\}$ indenter 
 planes, only slip traces aligned in $\langle110\rangle$ directions can be 
 produced through its interaction with dislocation gliding in $\{111\}$ 
 planes (Fig. \ref{fig:ExpResult}(a)).
 The grain with [101] orientation presents 
 a little higher hardness, the difference
 does not exceed 4\% (1.69 GPa vs 1.63 GPa). 
 The crystallographic aspects of the deformation process
 provide a comprehensive understanding
 of the material’s response to nanoindentation. 
 This combined approach of nanoindentation and EBSD imaging
 contributes valuable data to elucidate the intricate mechanics
 of plastic deformation in the NiFeCrCo alloy.


As discussed, the experimental nanoindentation test motivates the computational 
modeling to explore the plastic deformation mechanism in 
the NiFeCrCo CSA at different crystal orientations. 
The morphology of the indented samples is assessed by
calculating atomic displacements concerning the initial
frame in our MD simulations. 
In Fig. \ref{fig:ExpResult}a--b), slip traces and pile--up 
formations for the NiFeCrCo CSA are depicted at [001] and 
[011] orientations, reaching a maximum indentation
depth of 3 nm. The observed propagation of slip traces aligns
with the slip planes of the face-centered cubic (FCC) sample
at each crystal orientation. Notably, the 4--fold shapes
for [001] and [011] correspond to typical shapes exhibited by FCC metals 
\cite{VARILLAS2017431,KURPASKA2022110639}.
Building on our prior research \cite{FRYDRYCH2023104644}, 
we highlight the distinctive effect of the four elements
in the CSA alloy, forming a halo around the indenter tip
irrespective of crystal orientation. 
In contrast, single--element FCC metals typically result
in step--shaped pile-ups. This behavior is attributed to
the intricate chemistry of the CSA alloy.

\begin{figure}[t!]
    \centering
    \includegraphics[width=0.48\textwidth]{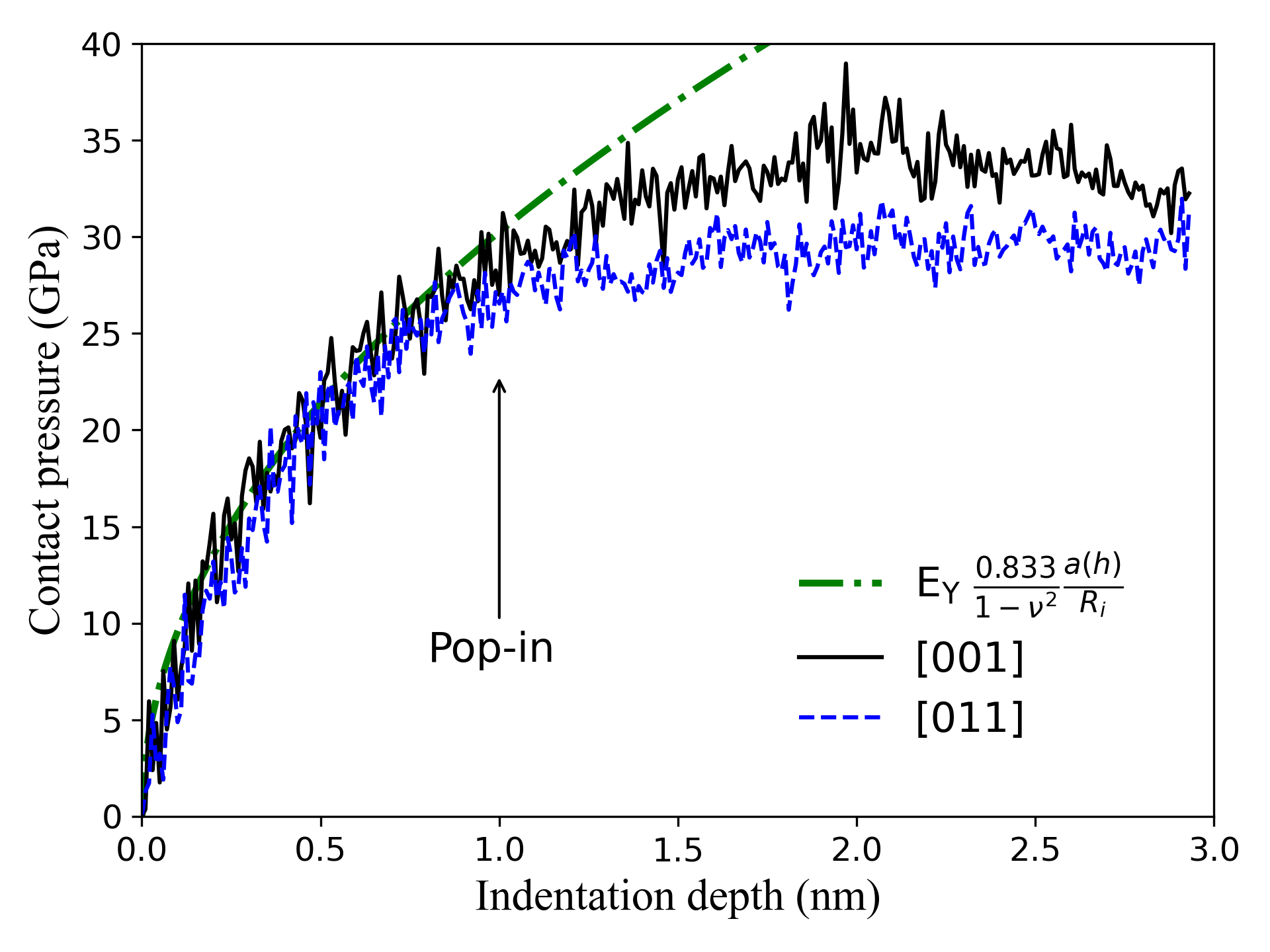}
    \caption{(Color online) Average contact pressure as a function of 
    indentation depth for the pristine NiFeCrCo sample at [001], [101], 
    and [111] orientations in (a), 
    with the pop--in event identified as a deviation from the
    linear scaling law. }
    \label{fig:avetabGAP}
\end{figure}

The implementation of the numerical modeling for nanoindentation tests is done 
in accordance to experimental conditions by tracking atomic strains and dislocation network dynamics. 
Fig. \ref{fig:avetabGAP} reports the mean contact pressure 
$p$, of the pristine CSA as a function of the nanoindentation depth 
which is obtained by
\cite{JavVarilla,REMINGTON2014378,PhysRevMaterials.7.043603}:
\begin{equation}
    p(h) = \frac{2\pi}{3E_Y} 
    \left[ 24P_{\rm ave}(h) \left( \frac{E_Y R_i}{1-\nu^2} \right)^2 \right]^{1/3},
\end{equation}
where $h$ is the indentation depth, $\nu$ is the Poisson's
ratio, and the average load is calculated as 
$P_{\rm ave}(h)=1/N \sum_i^N P_i(h)$ with $P_i(h)$ 
as the load from each MD simulation and $N=10$ the number
of indents.
During loading process, the contact radius is obtained with
the geometrical relationship as
\cite{PhysRevMaterials.7.043603}: 
\begin{equation}
\mathrm{a}(h) = \left[ 3P_{\rm ave}(h)R_i \frac{1-\nu^2}{8E_{\rm Y}} \right]^{1/3},
 \end{equation}
 these quantities provide an intrinsic measure of the surface 
 resistance to defect 
 nucleation~\cite{VARILLAS2017431,PhysRevMaterials.7.043603}
 and yield to a universal linear relationship between $p(h)/E_Y$
 and $a(h)/R_i$ given by 
 \begin{equation}
     \frac{p(h)}{E_Y} = \frac{0.844}{1-\nu^2} \frac{a(h)}{R_i}.
 \end{equation}
which is shown in the Fig \ref{fig:avetabGAP} by a green line. 
Hence, the pop-in event can be characterized as the departure of the 
contact pressure curve from the linear scaling law that describes the 
transition from the elastic to the plastic deformation region. It is 
noteworthy that the occurrence of this event is contingent upon the 
crystallographic orientation of the material where the critical load 
is higher for [001] than the one for [011] orientation. 

In order to analyze the influence of the crystal orientation on
the dislocation nucleation and evolution of the sample, we
visualize and quantify different types of dislocations nucleated
at different indentation depths by using the OVITO \cite{ovito}
software. 
This was done through the use of the Dislocation Extraction
Algorithm (DXA)~\cite{Stukowski_2012}; that extracts
dislocation structure and content from atomistic
microstructures. 
Thus, we categorized the dislocations into several dislocation
types according to their Burgers vectors as: ½$\langle110\rangle$ 
(Perfect), 1/6$\langle112\rangle$ (Shockley),
1/6$\langle110\rangle$ (Stair--rod), 1/3$\langle100\rangle$
(Hirth), 1/3$\rangle111\langle$ (Frank) noticing that
the nucleation of partial 1/6$\langle112\rangle$ Shockley
dislocations is dominant in the loading process regardless
of the crystal orientation due to the material's FCC structure. 
Thus, we compute the dislocation  density, $\rho$,
as a function of the depth as:
\begin{equation}
    \rho = \frac{N_D l_D}{V_D},
\end{equation}
where $N_D$ is the number of dislocation lines and loops measured during 
nanoindentation test; $l_D$ is the dislocation length of each type, 
and $V_D = 2\pi/3(R_{\rm pl}^3-h^3)$ is the volume of the plastic 
deformation region by using the approximation of a spherical 
plastic zone; where $R_{\rm pl}$ is the largest distance of 
a dislocation measured from the indentation displacement, 
considering a hemispherical geometry.

\begin{figure}[b!]
    \centering
    \includegraphics[width=0.48\textwidth]{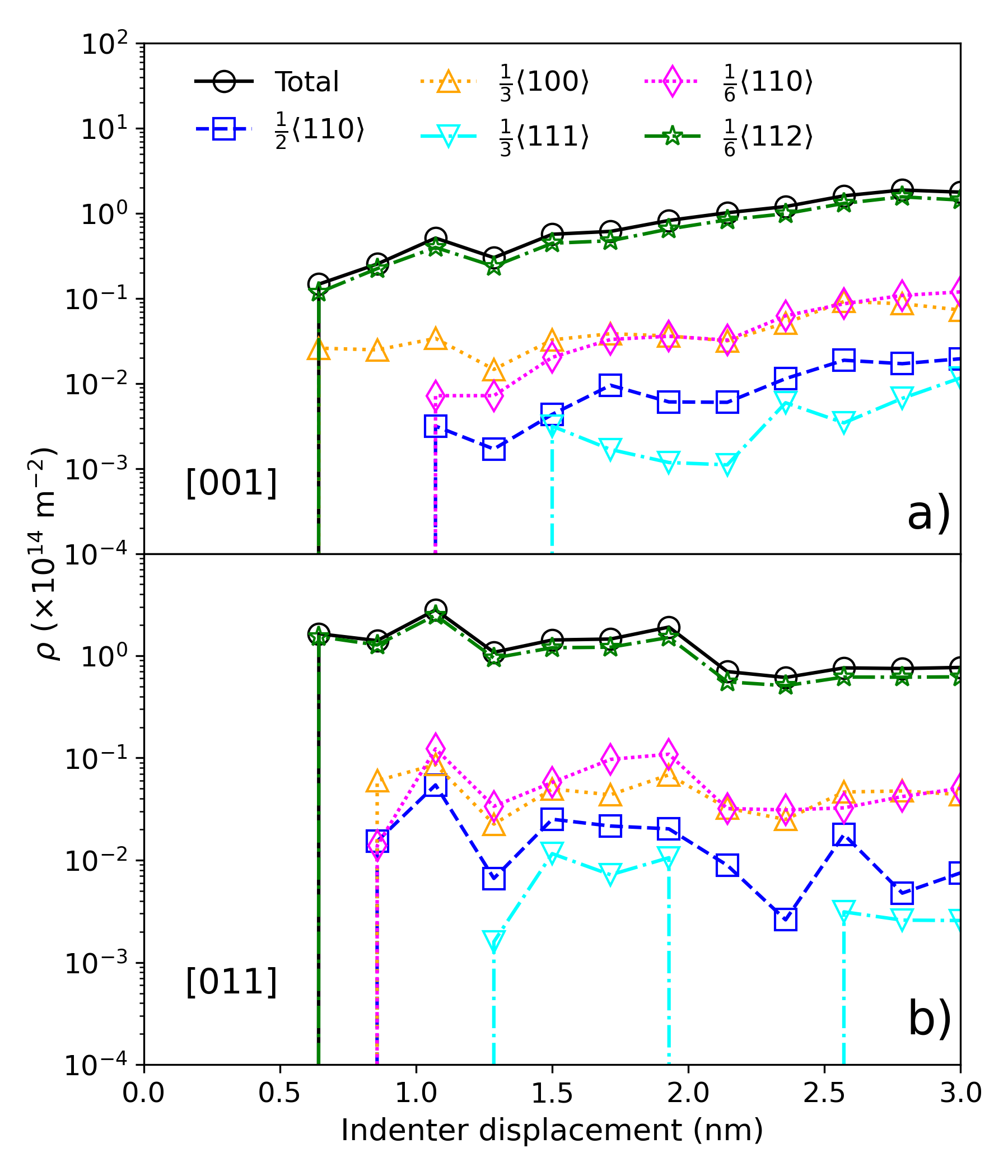}
    \caption{(Color online)  Average dislocation density plotted against 
    the indentation displacement for [100] and [101] crystal 
    orientations in (a) and (b), respectively. Notable 
    crystallographic effects are evident in dislocation nucleation. As 
    anticipated, Shockley dislocations predominantly nucleate and evolve 
    throughout the entire loading process, with their interactions giving 
    rise to stair--rod and Hirth junction dislocation nucleation.}
    \label{fig:DislocationDensity}
\end{figure}

In Fig. \ref{fig:DislocationDensity} we present the average of the
dislocation 
density as a function of the indentation depth for the [001] in a) 
and [011] in b) crystal orientation. 
It is observed that the Shockley dislocation is nucleated and evolved 
during the whole loading process, as expected. The interaction of different 
Shockley dislocations can lead to the nucleation of Hirth and Stair-rod dislocation as follows:
\begin{eqnarray}
    \frac{1}{3} [100] &=& \frac{1}{6} [1 2 \Bar{1}] 
    + \frac{1}{6}[1 \Bar{2} 1] \quad \rm{Hirth}, \\
    \frac{1}{6} [110] &=& \frac{1}{6} [\Bar{1} 2 1] + 
    \frac{1}{6} [2 \Bar{1} \Bar{1}] \quad \rm{Stair-rod},
\end{eqnarray}
and other symmetrical cases.
As the primary dislocation junctions manifest during the loading
process, this mechanism is recognized for its role in initiating
the nucleation of prismatic dislocation loops. 
This initiation is discerned as a decrease in the total
dislocation density at 1.1 nm across all orientations. 
Additionally, the nucleation of dislocations commences at
a depth of $0.75$ nm for [001] and [011] orientations. 
This discrepancy is attributed to the distinct
atomic arrangement within the unit cell of each orientation.
Moreover, it is noteworthy that the density of Schokley partials tends to rise for the $\{100\}$ indenter plane, in contrast to the slight decrease observed in the case of the $\{110\}$ indenter plane. This phenomenon may be partially explained by the strong interaction of these dislocations with the $\{110\}$ plane, leading to their annihilation, as evidenced by the slip traces shown Fig. \ref{fig:ExpResult}(b).

\begin{figure}[b!]
    \centering
    \includegraphics[width=0.48\textwidth]{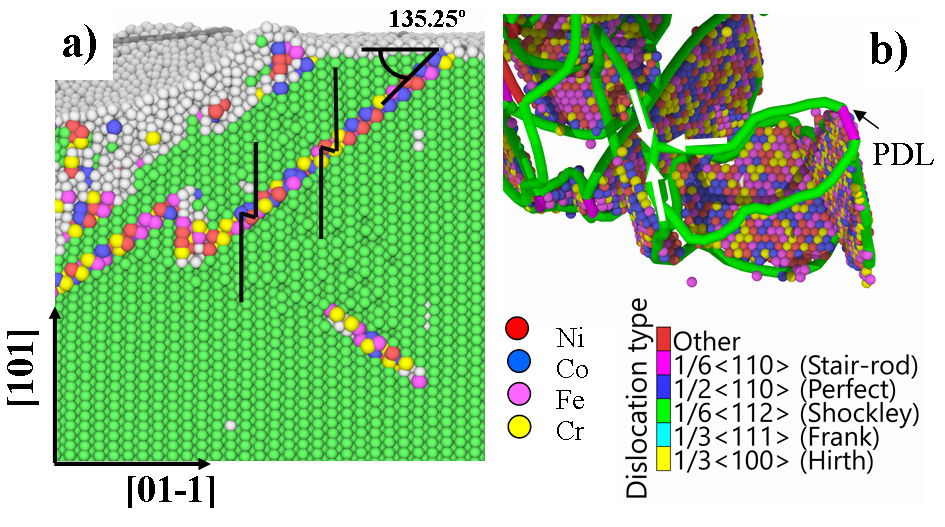}
    \caption{(Color online) Cross-sectional view of a nanotwin underneath 
    a [101] surface in (a); black lines indicate parent and twined 
    crystal orientations for enhanced visualization, illustrating the 
    well-known nanotwin mechanism in FCC metals. Formation of a prismatic 
    dislocation loop in the [101] orientation at an indentation 
    depth of 2 nm; white arrows indicate the dislocation junction. FCC 
    atoms are removed, and only atoms with HCP structure corresponding to 
    stacking fault planes are displayed for clarity. }
    \label{fig:twinning}
\end{figure}

At the maximum indentation depth, we observe parallels
between the NiFeCrCo CSA and those exhibited by single-element
face-centered cubic (FCC) metals, such as Ni 
\cite{VARILLAS2017431,KURPASKA2022110639}, as depicted in Fig. 
\ref{fig:twinning}. 
Figure \ref{fig:twinning}a) illustrates the formation of a
nano-twin beneath a [101] surface, with black lines denoting
parent and twinned crystal orientations. This observation aligns
well with the typical mechanisms observed in FCC metals. 
Importantly, our findings reveal that atomic ordering is not a 
prerequisite for this mechanism. 
Furthermore, twin nucleation in our sample is initiated under
the strain gradients and stresses imposed by the nanoindenter
tip. This involves the successive emission of leading partial 
dislocations in the vicinity of the indenter, resulting in the 
distinctive arrangement of parallel \{111\} twinned planes 
\cite{DAPHALAPURKAR2012277,KALIDINDI1998267}. Our 
simulations demonstrate that the twin boundaries are normal to the 
surface for (011) indentation and inclined at 135.25°. 
The twinning crystallography 
ensures that all $\langle011\rangle$ and $\langle112\rangle$
traces simultaneously lie at the indented plane and a specific
\{111\} habit plane, conforming to expectations for FCC metals.

In Fig. \ref{fig:twinning}b, we illustrate the prismatic dislocation
loop at the maximum indentation depth for the [101] crystal
orientation. Notably, the observation reveals that the junction
of Shockley partial dislocations leads to the final state of the 
formation of this loop, accompanied by stacking fault planes
identified as hexagonal close-packed (HCP) atoms. 
This identification is made through Polyhedral Template analysis
in Ovito. The atoms involved are color-coded based on their types, 
enabling an analysis of their random distribution maintained during
the loading process. Importantly, no specific ordering of atoms
was observed throughout this process.

Nevertheless, in comparison to pure Ni, all defects are confined to a 
narrow region beneath the surface, possibly attributed to the reduced 
dislocation velocity in HEA \cite{shen2021mobility}. That can be correlate 
to the typical hallow observed around the indenter tip a contrario to the 
larger step-shaped piles ups observed in pure Ni \cite{FRYDRYCH2023104644, 
alhafez2019nanoindentation,KURPASKA2022110639,Dominguez-Gutierrez_2022}.
In these alloys, dislocations exhibit a wavy morphology and encounter 
natural pinning points stemming from the variable chemical and energetic 
landscape surrounding the dislocation core. For the same reason twin 
boundary planes are expected to be more fragmented than in pure Ni \cite{alhafez2019nanoindentation}.

\subsection{Nanoindentation Modeling of defected NiFeCrCo CSA}

\begin{figure}[b!]
    \centering
    \includegraphics[width=0.48\textwidth]{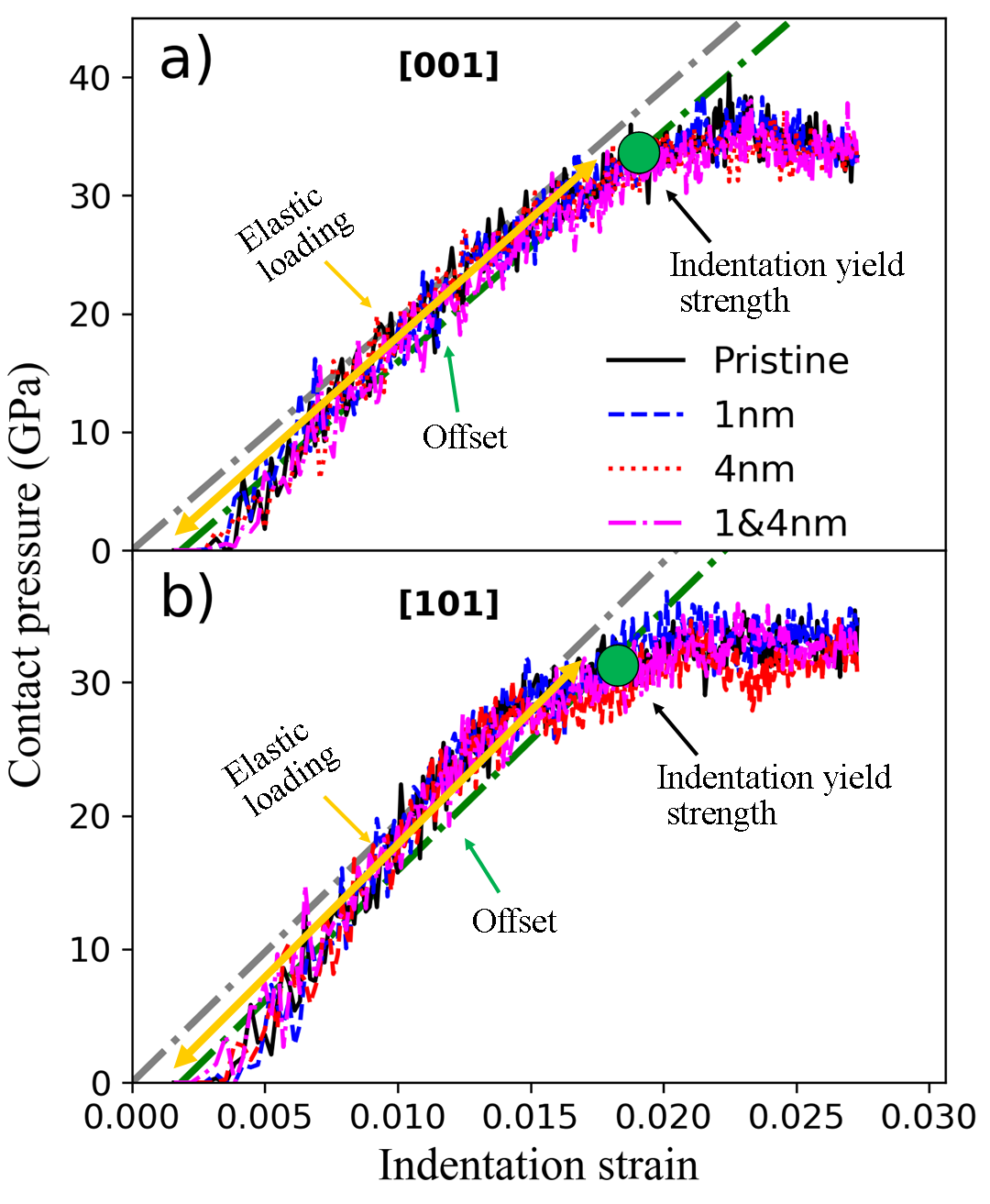}
    \caption{(Color online) Average nanoindentation contact pressure as a function of indentation strain, comparing it with the pristine case for [001] and [101] orientations. The pressure-strain curve is characterized by contrasting it with a linear scaling law (gray dashed line), where the offset in nanoindentation strain due to the indenter tip speed is indicated by a green line. The pop-in event, marking the indentation yield strength, is identified as a deviation from the scaling law, denoted by a green spot. The plastic region exhibits a stable pressure at higher nanoindentation strain values.}
    \label{fig:SSnano}
\end{figure}

The impact of the nanomechanical response in defected
NiFeCrCo CSA is examined through the analysis of nanoindentation
contact pressure as a function of indentation strain, comparing
it to the pristine case for both [001] and [101] orientations, as 
depicted in Fig. \ref{fig:SSnano}. 
The pressure--strain curve is characterized by contrasting it
with a linear scaling law to delineate the elastic loading phase.
Due to the speed of the indenter tip, there is an offset in
the nanoindentation strain, indicated by a green line. 
The occurrence of a pop--in event is identified as a deviation
of the pressure curve from the scaling law, marked by a green
spot, representing the indentation yield strength. 
Subsequently, the plastic region exhibits a stable pressure
at higher nanoindentation strain values.
Through the analysis of this material response during
the loading process, it is observed that materials with
preexisting Stacking Fault Tetrahedra (SFTs) experience
an increase in hardness values and their maximum shear stress 
\cite{PhysRevMaterials.7.043603}. 
SFT impede the glide of dislocations, requiring them to
find alternative paths to navigate around and pass through the SFT.

\begin{figure}[b!]
    \centering
    \includegraphics[width=0.48\textwidth]{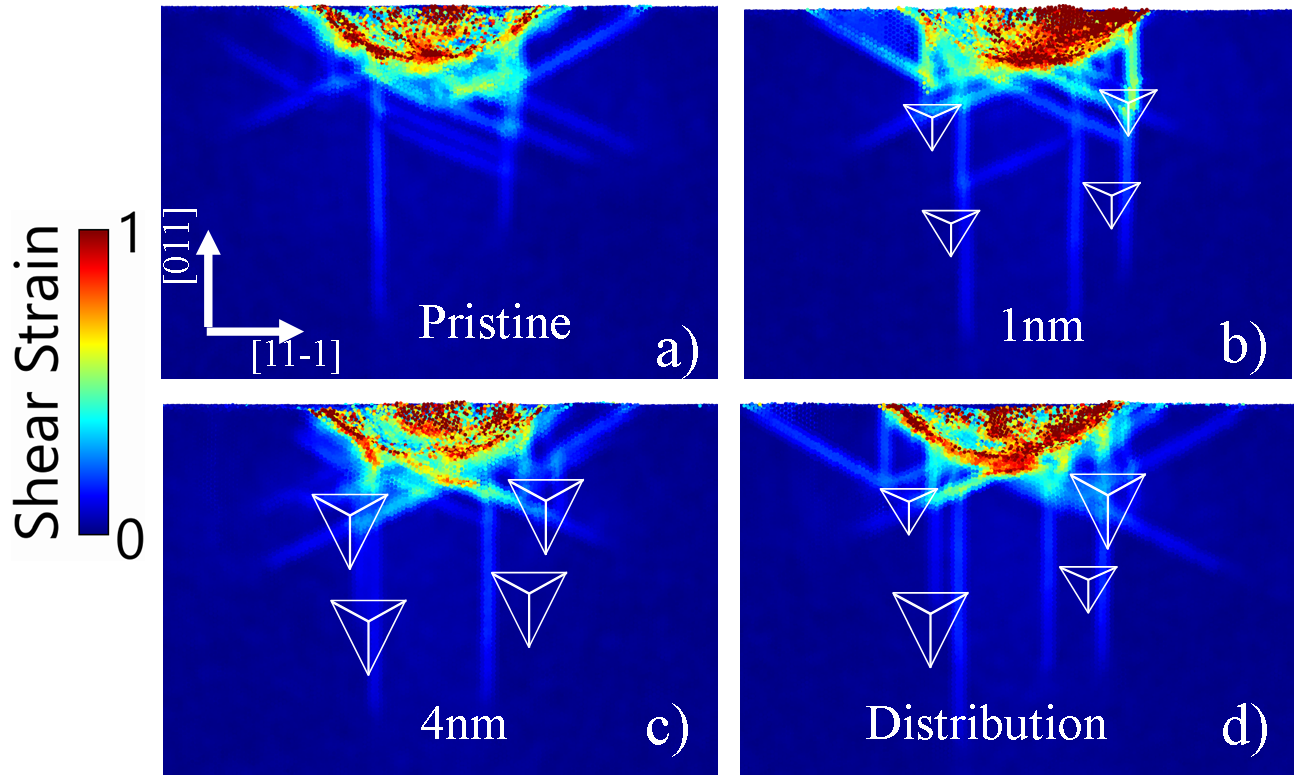}
    \caption{(Color online) Atomic shear strain mapping of selected MD 
    simulations of [101] NiFeCrCo alloy at the maximum indentation depth 
    for pristine samples in (a), with 1nm stacking fault tetrahedra 
    (SFTs) in (b), 4nm in (c), and SFTs at different sizes in (d). The 
    influence of defect materials is evident in the accumulation of 
    strain underneath the tip, affecting the propagation of dislocations. 
    }
    \label{fig:shearStrainMap}
\end{figure}

At the maximum indentation depth, the plastic deformation of the
NiFeCrCo CSA sample induced by the indenter tip's shear strain
becomes evident. In Fig. \ref{fig:shearStrainMap}, atomic
shear strain mappings are presented for the [011]
NiFeCrCo CSA sample under different conditions: pristine case
in (a), 1 nm of SFTs in (b), 4 nm of SFT in (c), 
and a combination of SFTs in (d). The sample is visually
enhanced by slicing it in half, and the width of the
"lamella" is halved for better visualization.
For the pristine case, maximum strain accumulates underneath
the indenter tip, and slip planes are delineated by an
increase in strain value relative to the FCC atoms
(colored in blue). In contrast, when SFTs are present, strain
accumulates on the surface for smaller SFTs, around the
positions of 4 nm-sized SFTs, and beneath the indenter tip.
This pattern persists for a mixed-size distribution of SFTs. 
Additionally, the triangular shape of the strained atoms varies
with the size of the SFTs. While the pristine case exhibits
a symmetric triangle, the presence of SFTs leads to asymmetric
and sharper triangles due to the pre-existing strain
fields around the defects. 
This effect, depending on the SFT size is challenging to understand
and may be link to the complex interactions between the SFT,
composed of stair rods dislocations and dislocations 
\cite{bacon2009dislocation}. 
This phenomenon influences the nucleation and evolution of dislocations, shear,
and prismatic dislocation loops.


\begin{figure}[b!]
    \centering
    \includegraphics[width=0.48\textwidth]{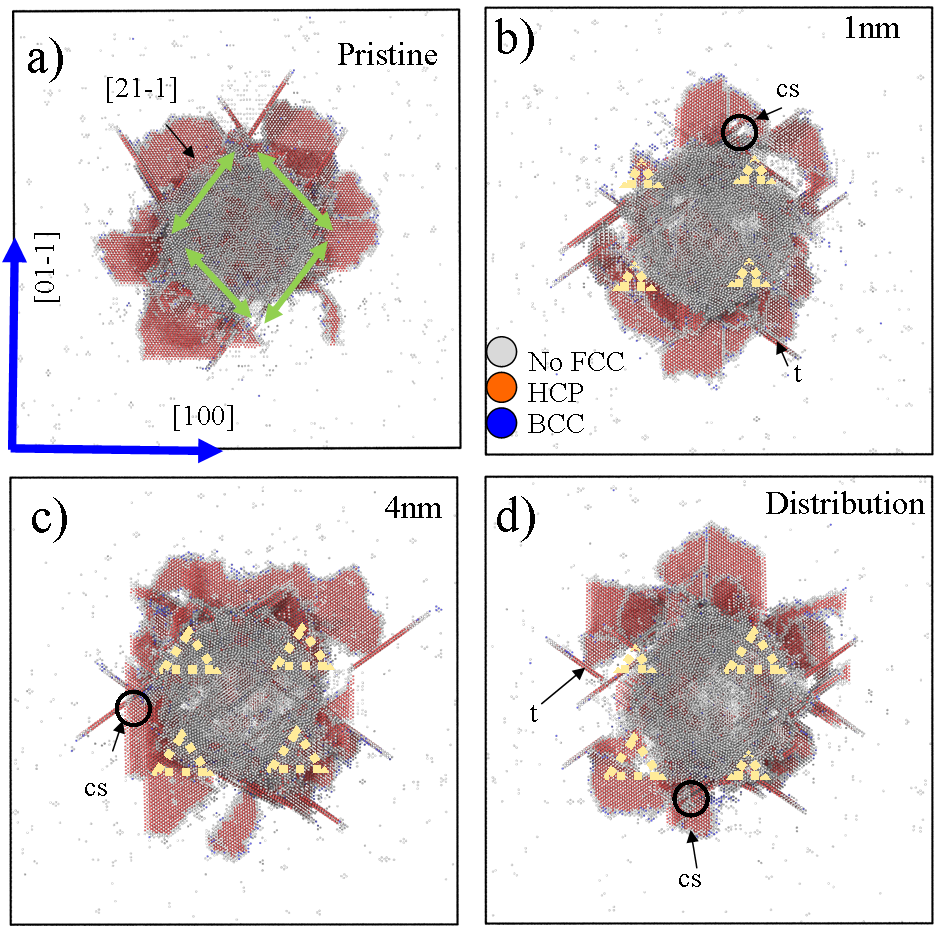}
    \caption{(Color online) Top view of the atomic structure of [101] 
    NiFeCrCo alloy for different samples (pristine, 1nm SFT, 4nm SFT, and 
    a combination of 1 and 4 nm SFT; indicated by dashed orange lines). 
    Slip planes are identified in the pristine case, while SFT promotes 
    the formation of cross-slip, highlighted by black circles. Nanotwins 
    are labeled with the letter "t."}
    \label{fig:TopView}
\end{figure}

In Fig. \ref{fig:TopView}, a top-view of the samples at the maximum 
indentation depth is presented, where the preexisting SFT vacancy defects 
are indicated by dashed yellow lines. It is observed that, for the 
pristine cases, the nanotwin and stacking fault planes propagate on the 
[011] slip planes and symmetric families. However, in the defected 
samples, these planes interact with the preexisting vacancy defects, 
forming a cross-slip in combination with a [001] plane, which is 
developed in the region between the vacancy defects.
This cross slip mechanisms, may arise from the interaction between
the SFT and the screw dislocation segments. 
Previous atomistic simulations show that the dislocation screw constricts 
when it reachs the SFT and then cross slipped \cite{bacon2009dislocation}. 
The interaction with edge dislocation segments leads to the formation of 
super-jogs, unstable that can lead to small vacancy clusters \cite{bacon2009dislocation}.

\begin{figure}[b!]
    \centering
    \includegraphics[width=0.48\textwidth]{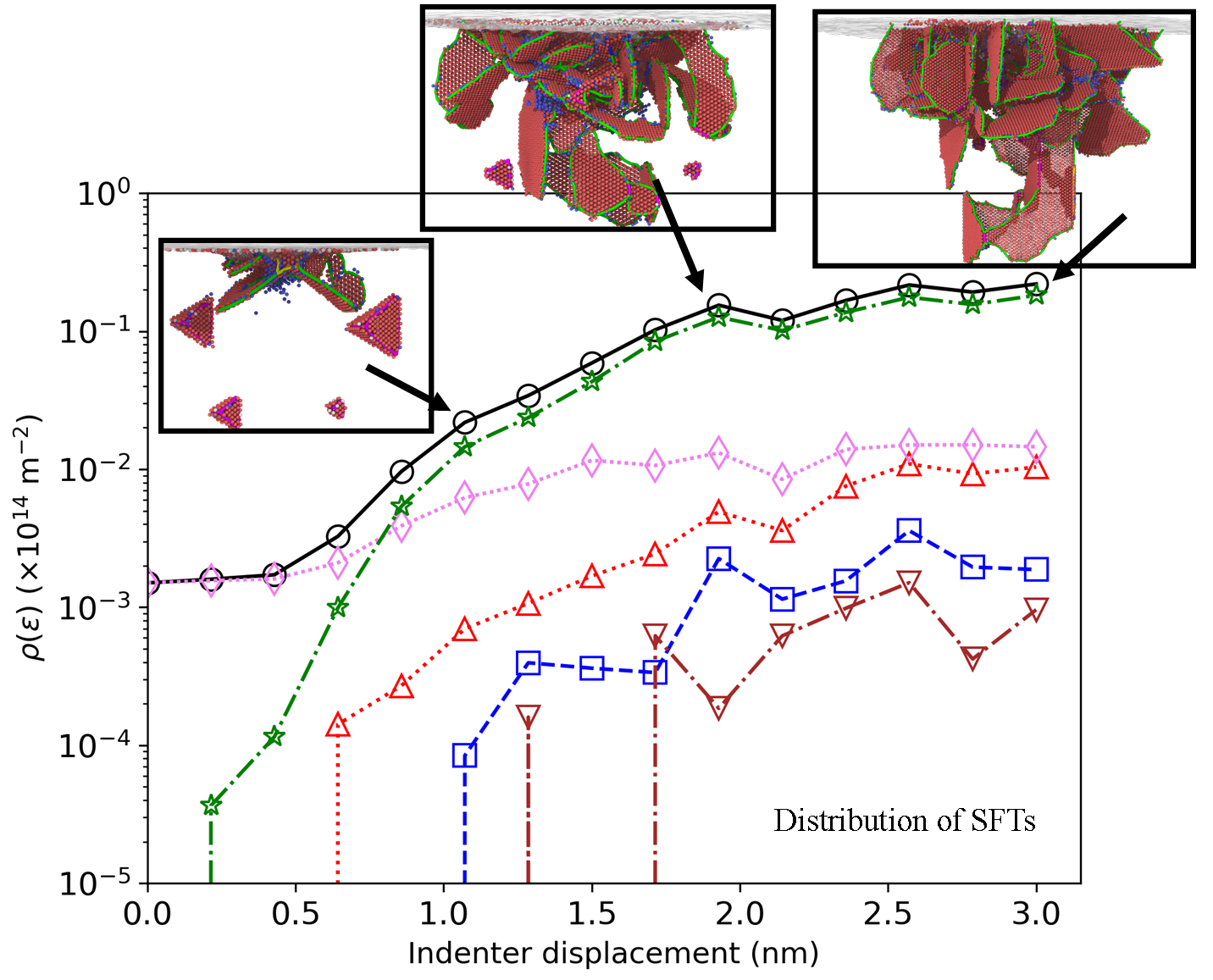}
    \caption{(Color online)Dislocation density plotted against indentation displacement for the NiFeCrMn CSA on the [011] orientation, featuring a size distribution for preexisting stacking fault tetrahedra (SFTs). The figure includes visualizations of dislocation nucleation and evolution at every nanometer of displacement, along with the formation of stacking planes. Color code for dislocation type and atomic structure follows the one use for Fig. \ref{fig:DislocationDensity} and \ref{fig:TopView}, respectively. }
    \label{fig:dislocDens_011SFT}
\end{figure}

To monitor the plastic deformation of the NiFeCrMn CSA during the
loading process, we examine the sample at the [011] orientation, 
considering a size distribution for preexisting SFTs and calculating
the dislocation density as a function of indentation displacement, 
as illustrated in Fig. \ref{fig:dislocDens_011SFT}.
Our observations reveal that SFTs are predominantly formed by 
Stair-rod type dislocations, and Shockley-like half-loops initiate 
nucleation during the early stages of nanoindentation loading, 
occurring within the 0 to 1.0 nm depth range, as depicted in the
inset of the figure. As the indentation progresses to greater depths, from 
1nm to 2nm, Shockley dislocations interact with the stair rods of the
SFT, absorbing them, while the stacking plane propagates due to
the stress applied by the indenter tip.
Upon reaching the maximum indentation depth, the larger-sized SFTs are 
already absorbed by the stacking fault planes induced by nanoindentation. 
Only smaller-sized SFTs survive the entire process, either by being
pushed downward or by not interacting with the stacking fault planes.

\subsection{The case of nanoindentation on the [111] orientation}

\begin{figure}[b!]
    \centering
    \includegraphics[width=0.48\textwidth]{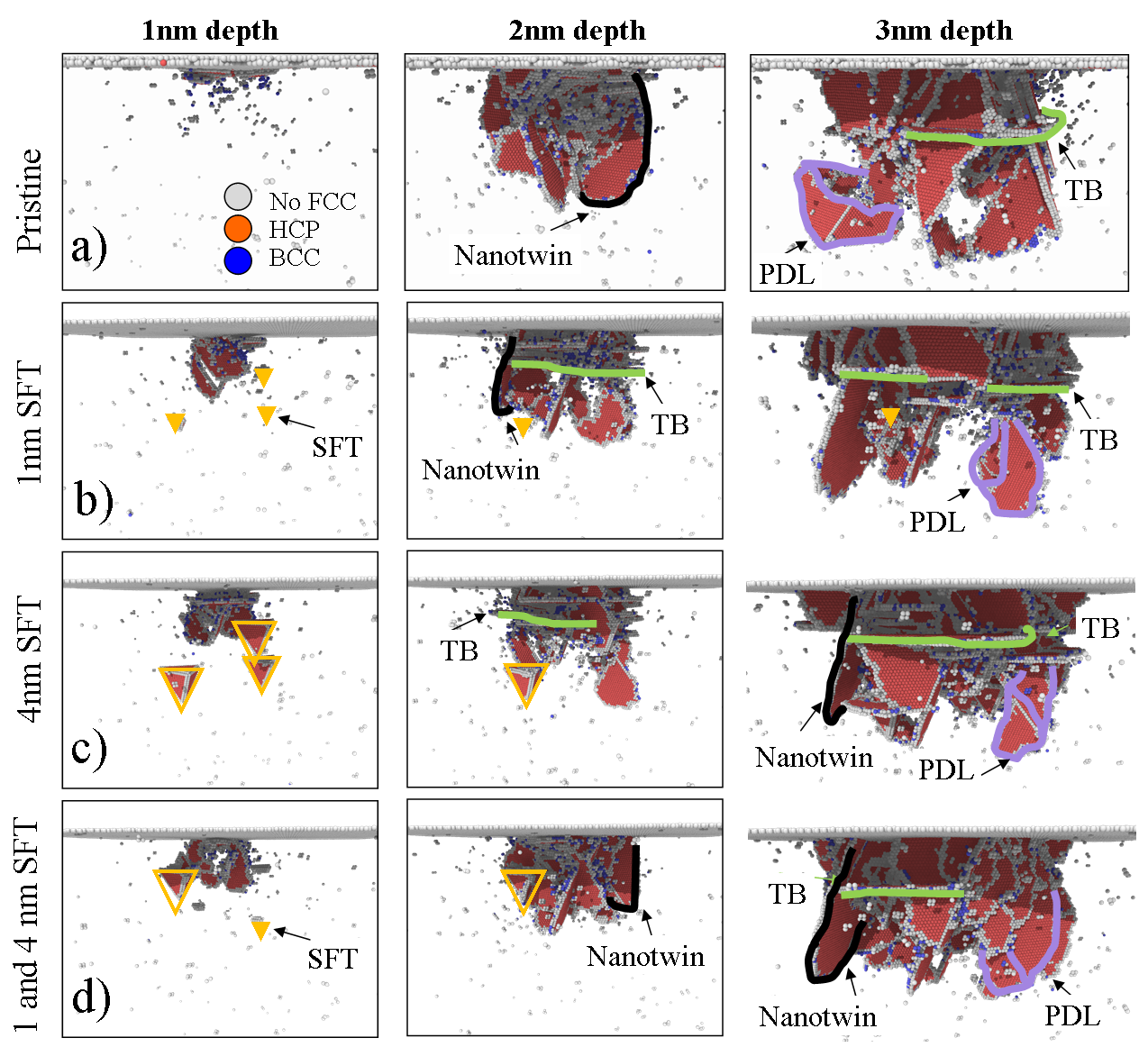}
    \caption{(Color online) Atomic structure visualization of [111] NiFeCrCo pristine and with preexisting stacking fault tetrahedra (SFTs) at different indentation depths: 1nm in (a), 2nm in (b), and 3nm in (c). FCC atoms are removed to illustrate the evolution of stacking fault planes as hexagonal close-packed (HCP) atoms, with nanotwins highlighted by black lines, twin boundaries depicted as green lines, and prismatic dislocation loops (PDLs) represented by violet lines. The influence of defected alloys is evident as strain at the surface alters the direction of plane propagation and induces PDL formation.}
    \label{fig:Load111}
\end{figure}

Although the experimental nanoindentation of the CSA on the [111] orientation 
is not often obtained, we analyze the mechanical response of the material 
on this specific orientation.
The Stacking Fault Tetrahedra (SFTs) are generated on the habit
\{111\} plane, aligning with the crystal orientation [111] in
the NiFeCrCo CSA. This creates an excess of pressure during
the loading process, attributed to the insertion of SFT vacancy
types highlighted by yellow lines.
Consequently, 
we conduct an analysis of the atomic structure evolution and
dislocation behavior in both pristine and defected samples
at various indentation depths.
In Fig. \ref{fig:Load111}, we illustrate the evolution of
stacking fault planes identified by hexagonal close-packed (HCP)
atoms in orange, as well as the propagation of dislocations
for the pristine case in (a) and SFT vacancy types in (b-d).
At 1 nm depth, the pristine case nucleates a small stacking fault
(SF) plane parallel to the \{111\} plane underneath the indenter
tip, while the strain field induced by SFT vacancy types
deforms this plane into a pair of half loops beneath the
indenter tip. At 2 nm depth, the pristine case shows a
nanotwin emerging from the side of the indenter tip,
while a twin boundary (TB) forms in samples with SFT
vacancy types. For smaller SFTs, the half loops do not
yet interact with the defects, but for a distribution
of SFT vacancy sizes, interactions with larger SFTs are observed.

At the maximum indentation depth of 3 nm, the pristine case has already 
initiated the nucleation of a complete prismatic dislocation loop (PDL) 
and a small twin boundary (TB). In contrast, the 1 nm Stacking Fault 
Tetrahedra (SFT) case exhibits the formation of a larger TB above the 
location of the SFT vacancies, and a small PDL is also nucleated in the 
direction where the SFT vacancies are absent. For larger 4 nm SFTs and a 
combination of SFT vacancies, the TB plane becomes larger, and PDLs begin 
to form in the same direction where there are no preexisting defects. 
These mechanisms have a significant impact on the nanomechanical response 
of the material, resulting in a load-displacement curve that is noisy and 
challenging to analyze for the defected cases.
The presence of SFTs at different depths induces interactions 
that influence dislocation nucleation and evolution. Consequently,
the resulting load-displacement curve in defected cases presents
a challenge for analysis due to its noisy response
due to excess of strain in the surface in close connection
to SFTs preexisting in the sample.
\section{\label{sec:concluding}Concluding remarks}

This study delves into the nanomechanical response of the
crystalline NiFeCrCo concentrated solid solution alloy (CSA)
through a comprehensive exploration employing both experimental
and computational approaches in nanoindentation tests. 
In the experimental realm, the production and characterization
of CSA leverage an arc melting technique. 
Subsequently, nanoindentation tests are conducted at room
temperature, allowing for the acquisition of surface morphology
through electron microscopy images.
In parallel, atomistic computational modeling captures the 
nanoindentation loading process, recording load--displacement curves, 
dislocation densities, and mapping atoms displacements. 
Notably, dislocation nucleation mechanisms are present with 
the prismatic loop formation profoundly influenced
by surface characteristics, stacking faults, and dislocation
glide energies during the nanoindentation loading process.
The experimental characterization of the NiFeCrCo CSA is in 
good agreement with MD simulations by showing similarities to 
the propagation of slip--traces and consecutive nucleation of 
prismatic dislocation loops.

In addition, the introduction of Stacking Fault Tetrahedra (SFT)
vacancy type defects emerges as a transformative factor in shaping
the nanomechanical behavior of the material.
Crystallographic orientations stand out as pivotal elements
influencing the evolution of dislocations and stacking fault planes 
during the loading process. Distinct behaviors observed in defected 
samples, including altered dislocation propagation 
and stacking fault planes, underscore the profound impact of 
crystallographic factors on the material's mechanical response.

Furthermore, the existence of preexisting SFT vacancy defects
triggers interactions with dislocation planes, giving rise
to phenomena such as cross-slip and the formation of additional
slip planes, notably [101]. The emergence of twin boundaries, prismatic 
dislocation loops, and modifications to crystallographic families 
collectively illuminate the intricate interplay between crystallographic 
factors and defect structures. This interplay significantly influences 
the material's mechanical response during nanoindentation.

\section*{Acknowledgements}
Research was funded through the European Union Horizon 2020 research and innovation program under Grant Agreement No. 857470 and from the European Regional Development Fund under the program of the Foundation for Polish Science International Research Agenda PLUS, grant No. MAB PLUS/2018/8, and the initiative of the Ministry of Science and Higher Education 'Support for the activities of Centers of Excellence established in Poland under the Horizon 2020 program' under agreement No. MEiN/2023/DIR/3795.
We gratefully acknowledge enlightening discussions with Xavier Feaugas. We would like to express our gratitude to NOETHER computing facilities at La Rochelle University and MCIA (Mésocentre de Calcul Intensif Atlantique). Additionally, we extend our gratitude to GENCI - (CINES/CCRT), under Grant number A0110913037.

\section*{References}
\bibliography{bibliography}

\begin{thebibliography}{52}%
\makeatletter
\providecommand \@ifxundefined [1]{%
 \@ifx{#1\undefined}
}%
\providecommand \@ifnum [1]{%
 \ifnum #1\expandafter \@firstoftwo
 \else \expandafter \@secondoftwo
 \fi
}%
\providecommand \@ifx [1]{%
 \ifx #1\expandafter \@firstoftwo
 \else \expandafter \@secondoftwo
 \fi
}%
\providecommand \natexlab [1]{#1}%
\providecommand \enquote  [1]{``#1''}%
\providecommand \bibnamefont  [1]{#1}%
\providecommand \bibfnamefont [1]{#1}%
\providecommand \citenamefont [1]{#1}%
\providecommand \href@noop [0]{\@secondoftwo}%
\providecommand \href [0]{\begingroup \@sanitize@url \@href}%
\providecommand \@href[1]{\@@startlink{#1}\@@href}%
\providecommand \@@href[1]{\endgroup#1\@@endlink}%
\providecommand \@sanitize@url [0]{\catcode `\\12\catcode `\$12\catcode `\&12\catcode `\#12\catcode `\^12\catcode `\_12\catcode `\%12\relax}%
\providecommand \@@startlink[1]{}%
\providecommand \@@endlink[0]{}%
\providecommand \url  [0]{\begingroup\@sanitize@url \@url }%
\providecommand \@url [1]{\endgroup\@href {#1}{\urlprefix }}%
\providecommand \urlprefix  [0]{URL }%
\providecommand \Eprint [0]{\href }%
\providecommand \doibase [0]{http://dx.doi.org/}%
\providecommand \selectlanguage [0]{\@gobble}%
\providecommand \bibinfo  [0]{\@secondoftwo}%
\providecommand \bibfield  [0]{\@secondoftwo}%
\providecommand \translation [1]{[#1]}%
\providecommand \BibitemOpen [0]{}%
\providecommand \bibitemStop [0]{}%
\providecommand \bibitemNoStop [0]{.\EOS\space}%
\providecommand \EOS [0]{\spacefactor3000\relax}%
\providecommand \BibitemShut  [1]{\csname bibitem#1\endcsname}%
\let\auto@bib@innerbib\@empty
\bibitem [{\citenamefont {Wagner}\ \emph {et~al.}(2022)\citenamefont {Wagner}, \citenamefont {Ferrari}, \citenamefont {Schreuer}, \citenamefont {CouziniĂ©}, \citenamefont {Ikeda}, \citenamefont {KĂ¶rmann}, \citenamefont {Eggeler}, \citenamefont {George},\ and\ \citenamefont {Laplanche}}]{WAGNER2022117693}%
  \BibitemOpen
  \bibfield  {author} {\bibinfo {author} {\bibfnamefont {C.}~\bibnamefont {Wagner}}, \bibinfo {author} {\bibfnamefont {A.}~\bibnamefont {Ferrari}}, \bibinfo {author} {\bibfnamefont {J.}~\bibnamefont {Schreuer}}, \bibinfo {author} {\bibfnamefont {J.-P.}\ \bibnamefont {CouziniĂ©}}, \bibinfo {author} {\bibfnamefont {Y.}~\bibnamefont {Ikeda}}, \bibinfo {author} {\bibfnamefont {F.}~\bibnamefont {KĂ¶rmann}}, \bibinfo {author} {\bibfnamefont {G.}~\bibnamefont {Eggeler}}, \bibinfo {author} {\bibfnamefont {E.~P.}\ \bibnamefont {George}}, \ and\ \bibinfo {author} {\bibfnamefont {G.}~\bibnamefont {Laplanche}},\ }\href {\doibase https://doi.org/10.1016/j.actamat.2022.117693} {\bibfield  {journal} {\bibinfo  {journal} {Acta Materialia}\ }\textbf {\bibinfo {volume} {227}},\ \bibinfo {pages} {117693} (\bibinfo {year} {2022})}\BibitemShut {NoStop}%
\bibitem [{\citenamefont {Li}\ and\ \citenamefont {et~al.}(2017)}]{li2017interstitial}%
  \BibitemOpen
  \bibfield  {author} {\bibinfo {author} {\bibfnamefont {Z.}~\bibnamefont {Li}}\ and\ \bibinfo {author} {\bibnamefont {et~al.}},\ }\href {\doibase 10.1038/srep40704} {\bibfield  {journal} {\bibinfo  {journal} {Sci. Rep.}\ }\textbf {\bibinfo {volume} {7}},\ \bibinfo {pages} {40704} (\bibinfo {year} {2017})}\BibitemShut {NoStop}%
\bibitem [{\citenamefont {Huo}\ \emph {et~al.}(2017)\citenamefont {Huo}, \citenamefont {Fang}, \citenamefont {Zhou}, \citenamefont {Xie}, \citenamefont {Shang},\ and\ \citenamefont {Jiang}}]{HUO2017125}%
  \BibitemOpen
  \bibfield  {author} {\bibinfo {author} {\bibfnamefont {W.}~\bibnamefont {Huo}}, \bibinfo {author} {\bibfnamefont {F.}~\bibnamefont {Fang}}, \bibinfo {author} {\bibfnamefont {H.}~\bibnamefont {Zhou}}, \bibinfo {author} {\bibfnamefont {Z.}~\bibnamefont {Xie}}, \bibinfo {author} {\bibfnamefont {J.}~\bibnamefont {Shang}}, \ and\ \bibinfo {author} {\bibfnamefont {J.}~\bibnamefont {Jiang}},\ }\href {\doibase https://doi.org/10.1016/j.scriptamat.2017.08.006} {\bibfield  {journal} {\bibinfo  {journal} {Scripta Materialia}\ }\textbf {\bibinfo {volume} {141}},\ \bibinfo {pages} {125} (\bibinfo {year} {2017})}\BibitemShut {NoStop}%
\bibitem [{\citenamefont {Shahmir}\ \emph {et~al.}(2017)\citenamefont {Shahmir}, \citenamefont {He}, \citenamefont {Lu}, \citenamefont {Kawasaki},\ and\ \citenamefont {Langdon}}]{SHAHMIR2017342}%
  \BibitemOpen
  \bibfield  {author} {\bibinfo {author} {\bibfnamefont {H.}~\bibnamefont {Shahmir}}, \bibinfo {author} {\bibfnamefont {J.}~\bibnamefont {He}}, \bibinfo {author} {\bibfnamefont {Z.}~\bibnamefont {Lu}}, \bibinfo {author} {\bibfnamefont {M.}~\bibnamefont {Kawasaki}}, \ and\ \bibinfo {author} {\bibfnamefont {T.~G.}\ \bibnamefont {Langdon}},\ }\href {\doibase https://doi.org/10.1016/j.msea.2017.01.016} {\bibfield  {journal} {\bibinfo  {journal} {Materials Science and Engineering: A}\ }\textbf {\bibinfo {volume} {685}},\ \bibinfo {pages} {342} (\bibinfo {year} {2017})}\BibitemShut {NoStop}%
\bibitem [{\citenamefont {Cichocki}\ \emph {et~al.}(2022)\citenamefont {Cichocki}, \citenamefont {Bała}, \citenamefont {Kozieł},\ and\ \citenamefont {et~al.}}]{cichocki2022effect}%
  \BibitemOpen
  \bibfield  {author} {\bibinfo {author} {\bibfnamefont {K.}~\bibnamefont {Cichocki}}, \bibinfo {author} {\bibfnamefont {P.}~\bibnamefont {Bała}}, \bibinfo {author} {\bibfnamefont {T.}~\bibnamefont {Kozieł}}, \ and\ \bibinfo {author} {\bibnamefont {et~al.}},\ }\href {\doibase 10.1007/s11661-022-06629-x} {\bibfield  {journal} {\bibinfo  {journal} {Metall Mater Trans A}\ }\textbf {\bibinfo {volume} {53}},\ \bibinfo {pages} {1749–1760} (\bibinfo {year} {2022})}\BibitemShut {NoStop}%
\bibitem [{\citenamefont {Frydrych}\ \emph {et~al.}(2021)\citenamefont {Frydrych}, \citenamefont {Karimi}, \citenamefont {Pecelerowicz}, \citenamefont {Alvarez}, \citenamefont {Dominguez-GutiĂ©rrez}, \citenamefont {Rovaris},\ and\ \citenamefont {Papanikolaou}}]{ma14195764}%
  \BibitemOpen
  \bibfield  {author} {\bibinfo {author} {\bibfnamefont {K.}~\bibnamefont {Frydrych}}, \bibinfo {author} {\bibfnamefont {K.}~\bibnamefont {Karimi}}, \bibinfo {author} {\bibfnamefont {M.}~\bibnamefont {Pecelerowicz}}, \bibinfo {author} {\bibfnamefont {R.}~\bibnamefont {Alvarez}}, \bibinfo {author} {\bibfnamefont {F.~J.}\ \bibnamefont {Dominguez-GutiĂ©rrez}}, \bibinfo {author} {\bibfnamefont {F.}~\bibnamefont {Rovaris}}, \ and\ \bibinfo {author} {\bibfnamefont {S.}~\bibnamefont {Papanikolaou}},\ }\href@noop {} {\bibfield  {journal} {\bibinfo  {journal} {Materials}\ }\textbf {\bibinfo {volume} {14}} (\bibinfo {year} {2021})}\BibitemShut {NoStop}%
\bibitem [{\citenamefont {George}\ \emph {et~al.}(2020)\citenamefont {George}, \citenamefont {Curtin},\ and\ \citenamefont {Tasan}}]{GEORGE2020435}%
  \BibitemOpen
  \bibfield  {author} {\bibinfo {author} {\bibfnamefont {E.}~\bibnamefont {George}}, \bibinfo {author} {\bibfnamefont {W.}~\bibnamefont {Curtin}}, \ and\ \bibinfo {author} {\bibfnamefont {C.}~\bibnamefont {Tasan}},\ }\href {\doibase https://doi.org/10.1016/j.actamat.2019.12.015} {\bibfield  {journal} {\bibinfo  {journal} {Acta Materialia}\ }\textbf {\bibinfo {volume} {188}},\ \bibinfo {pages} {435} (\bibinfo {year} {2020})}\BibitemShut {NoStop}%
\bibitem [{\citenamefont {He}\ \emph {et~al.}(2017)\citenamefont {He} \emph {et~al.}}]{HeNanoindentation}%
  \BibitemOpen
  \bibfield  {author} {\bibinfo {author} {\bibfnamefont {Q.~F.}\ \bibnamefont {He}} \emph {et~al.},\ }\href@noop {} {\bibfield  {journal} {\bibinfo  {journal} {Materials Research Letters}\ }\textbf {\bibinfo {volume} {5}},\ \bibinfo {pages} {300} (\bibinfo {year} {2017})}\BibitemShut {NoStop}%
\bibitem [{\citenamefont {MusicĂł}\ \emph {et~al.}(2020)\citenamefont {MusicĂł}, \citenamefont {Gilbert}, \citenamefont {Ward}, \citenamefont {Page}, \citenamefont {George}, \citenamefont {Yan}, \citenamefont {Mandrus},\ and\ \citenamefont {Keppens}}]{MusicalAPL}%
  \BibitemOpen
  \bibfield  {author} {\bibinfo {author} {\bibfnamefont {B.~L.}\ \bibnamefont {MusicĂł}}, \bibinfo {author} {\bibfnamefont {D.}~\bibnamefont {Gilbert}}, \bibinfo {author} {\bibfnamefont {T.~Z.}\ \bibnamefont {Ward}}, \bibinfo {author} {\bibfnamefont {K.}~\bibnamefont {Page}}, \bibinfo {author} {\bibfnamefont {E.}~\bibnamefont {George}}, \bibinfo {author} {\bibfnamefont {J.}~\bibnamefont {Yan}}, \bibinfo {author} {\bibfnamefont {D.}~\bibnamefont {Mandrus}}, \ and\ \bibinfo {author} {\bibfnamefont {V.}~\bibnamefont {Keppens}},\ }\href@noop {} {\bibfield  {journal} {\bibinfo  {journal} {APL Materials}\ }\textbf {\bibinfo {volume} {8}},\ \bibinfo {pages} {040912} (\bibinfo {year} {2020})}\BibitemShut {NoStop}%
\bibitem [{\citenamefont {Yeh}\ \emph {et~al.}()\citenamefont {Yeh}, \citenamefont {Chen}, \citenamefont {Lin}, \citenamefont {Gan}, \citenamefont {Chin}, \citenamefont {Shun}, \citenamefont {Tsau},\ and\ \citenamefont {Chang}}]{YehHEA}%
  \BibitemOpen
  \bibfield  {author} {\bibinfo {author} {\bibfnamefont {J.-W.}\ \bibnamefont {Yeh}}, \bibinfo {author} {\bibfnamefont {S.-K.}\ \bibnamefont {Chen}}, \bibinfo {author} {\bibfnamefont {S.-J.}\ \bibnamefont {Lin}}, \bibinfo {author} {\bibfnamefont {J.-Y.}\ \bibnamefont {Gan}}, \bibinfo {author} {\bibfnamefont {T.-S.}\ \bibnamefont {Chin}}, \bibinfo {author} {\bibfnamefont {T.-T.}\ \bibnamefont {Shun}}, \bibinfo {author} {\bibfnamefont {C.-H.}\ \bibnamefont {Tsau}}, \ and\ \bibinfo {author} {\bibfnamefont {S.-Y.}\ \bibnamefont {Chang}},\ }\href@noop {} {\bibfield  {journal} {\bibinfo  {journal} {Advanced Engineering Materials}\ }\textbf {\bibinfo {volume} {6}},\ \bibinfo {pages} {299}}\BibitemShut {NoStop}%
\bibitem [{\citenamefont {Babilas}\ \emph {et~al.}(2023)\citenamefont {Babilas}, \citenamefont {Mlynarek-Lak}, \citenamefont {Radol}, \citenamefont {lolski} \emph {et~al.}}]{BABILAS2023170839}%
  \BibitemOpen
  \bibfield  {author} {\bibinfo {author} {\bibfnamefont {R.}~\bibnamefont {Babilas}}, \bibinfo {author} {\bibfnamefont {K.}~\bibnamefont {Mlynarek-Lak}}, \bibinfo {author} {\bibfnamefont {A.}~\bibnamefont {Radol}}, \bibinfo {author} {\bibfnamefont {W.}~\bibnamefont {lolski}},  \emph {et~al.},\ }\href {\doibase https://doi.org/10.1016/j.jallcom.2023.170839} {\bibfield  {journal} {\bibinfo  {journal} {Journal of Alloys and Compounds}\ }\textbf {\bibinfo {volume} {960}},\ \bibinfo {pages} {170839} (\bibinfo {year} {2023})}\BibitemShut {NoStop}%
\bibitem [{\citenamefont {Miracle}\ and\ \citenamefont {Senkov}(2017)}]{MIRACLE2017448}%
  \BibitemOpen
  \bibfield  {author} {\bibinfo {author} {\bibfnamefont {D.}~\bibnamefont {Miracle}}\ and\ \bibinfo {author} {\bibfnamefont {O.}~\bibnamefont {Senkov}},\ }\href {\doibase https://doi.org/10.1016/j.actamat.2016.08.081} {\bibfield  {journal} {\bibinfo  {journal} {Acta Materialia}\ }\textbf {\bibinfo {volume} {122}},\ \bibinfo {pages} {448} (\bibinfo {year} {2017})}\BibitemShut {NoStop}%
\bibitem [{\citenamefont {Huo}\ \emph {et~al.}(2019)\citenamefont {Huo}, \citenamefont {Fang}, \citenamefont {Liu}, \citenamefont {Tan}, \citenamefont {Xie},\ and\ \citenamefont {Jiang}}]{WenyiAPL}%
  \BibitemOpen
  \bibfield  {author} {\bibinfo {author} {\bibfnamefont {W.}~\bibnamefont {Huo}}, \bibinfo {author} {\bibfnamefont {F.}~\bibnamefont {Fang}}, \bibinfo {author} {\bibfnamefont {X.}~\bibnamefont {Liu}}, \bibinfo {author} {\bibfnamefont {S.}~\bibnamefont {Tan}}, \bibinfo {author} {\bibfnamefont {Z.}~\bibnamefont {Xie}}, \ and\ \bibinfo {author} {\bibfnamefont {J.}~\bibnamefont {Jiang}},\ }\href@noop {} {\bibfield  {journal} {\bibinfo  {journal} {Applied Physics Letters}\ }\textbf {\bibinfo {volume} {114}},\ \bibinfo {pages} {101904} (\bibinfo {year} {2019})}\BibitemShut {NoStop}%
\bibitem [{\citenamefont {Olejarz}\ \emph {et~al.}(2023)\citenamefont {Olejarz}, \citenamefont {Huo}, \citenamefont {Zielinski}, \citenamefont {Diduszko}, \citenamefont {Wyszkowska}, \citenamefont {Kosinska}, \citenamefont {Kalita}, \citenamefont {Janwik}, \citenamefont {Chmielewski}, \citenamefont {Fang},\ and\ \citenamefont {Kurpaska}}]{OLEJARZ2023168196}%
  \BibitemOpen
  \bibfield  {author} {\bibinfo {author} {\bibfnamefont {A.}~\bibnamefont {Olejarz}}, \bibinfo {author} {\bibfnamefont {W.}~\bibnamefont {Huo}}, \bibinfo {author} {\bibfnamefont {M.}~\bibnamefont {Zielinski}}, \bibinfo {author} {\bibfnamefont {R.}~\bibnamefont {Diduszko}}, \bibinfo {author} {\bibfnamefont {E.}~\bibnamefont {Wyszkowska}}, \bibinfo {author} {\bibfnamefont {A.}~\bibnamefont {Kosinska}}, \bibinfo {author} {\bibfnamefont {D.}~\bibnamefont {Kalita}}, \bibinfo {author} {\bibfnamefont {I.}~\bibnamefont {Janwik}}, \bibinfo {author} {\bibfnamefont {M.}~\bibnamefont {Chmielewski}}, \bibinfo {author} {\bibfnamefont {F.}~\bibnamefont {Fang}}, \ and\ \bibinfo {author} {\bibfnamefont {L.}~\bibnamefont {Kurpaska}},\ }\href {\doibase https://doi.org/10.1016/j.jallcom.2022.168196} {\bibfield  {journal} {\bibinfo  {journal} {Journal of Alloys and Compounds}\ }\textbf {\bibinfo {volume} {938}},\ \bibinfo {pages} {168196} (\bibinfo {year} {2023})}\BibitemShut {NoStop}%
\bibitem [{\citenamefont {Karimi}\ \emph {et~al.}(2023)\citenamefont {Karimi}, \citenamefont {Salmenjoki}, \citenamefont {Mulewska}, \citenamefont {Kurpaska}, \citenamefont {Kosinska}, \citenamefont {Alava},\ and\ \citenamefont {Papanikolaou}}]{KARIMI2023115559}%
  \BibitemOpen
  \bibfield  {author} {\bibinfo {author} {\bibfnamefont {K.}~\bibnamefont {Karimi}}, \bibinfo {author} {\bibfnamefont {H.}~\bibnamefont {Salmenjoki}}, \bibinfo {author} {\bibfnamefont {K.}~\bibnamefont {Mulewska}}, \bibinfo {author} {\bibfnamefont {L.}~\bibnamefont {Kurpaska}}, \bibinfo {author} {\bibfnamefont {A.}~\bibnamefont {Kosinska}}, \bibinfo {author} {\bibfnamefont {M.~J.}\ \bibnamefont {Alava}}, \ and\ \bibinfo {author} {\bibfnamefont {S.}~\bibnamefont {Papanikolaou}},\ }\href {\doibase https://doi.org/10.1016/j.scriptamat.2023.115559} {\bibfield  {journal} {\bibinfo  {journal} {Scripta Materialia}\ }\textbf {\bibinfo {volume} {234}},\ \bibinfo {pages} {115559} (\bibinfo {year} {2023})}\BibitemShut {NoStop}%
\bibitem [{\citenamefont {Kurpaska}\ \emph {et~al.}(2022)\citenamefont {Kurpaska}, \citenamefont {Dominguez-Gutierrez}, \citenamefont {Zhang}, \citenamefont {Mulewska}, \citenamefont {Bei}, \citenamefont {Weber}, \citenamefont {Kosinska}, \citenamefont {Chrominski}, \citenamefont {Jozwik}, \citenamefont {Alvarez-Donado}, \citenamefont {Papanikolaou}, \citenamefont {Jagielski},\ and\ \citenamefont {Alava}}]{KURPASKA2022110639}%
  \BibitemOpen
  \bibfield  {author} {\bibinfo {author} {\bibfnamefont {L.}~\bibnamefont {Kurpaska}}, \bibinfo {author} {\bibfnamefont {F.}~\bibnamefont {Dominguez-Gutierrez}}, \bibinfo {author} {\bibfnamefont {Y.}~\bibnamefont {Zhang}}, \bibinfo {author} {\bibfnamefont {K.}~\bibnamefont {Mulewska}}, \bibinfo {author} {\bibfnamefont {H.}~\bibnamefont {Bei}}, \bibinfo {author} {\bibfnamefont {W.}~\bibnamefont {Weber}}, \bibinfo {author} {\bibfnamefont {A.}~\bibnamefont {Kosinska}}, \bibinfo {author} {\bibfnamefont {W.}~\bibnamefont {Chrominski}}, \bibinfo {author} {\bibfnamefont {I.}~\bibnamefont {Jozwik}}, \bibinfo {author} {\bibfnamefont {R.}~\bibnamefont {Alvarez-Donado}}, \bibinfo {author} {\bibfnamefont {S.}~\bibnamefont {Papanikolaou}}, \bibinfo {author} {\bibfnamefont {J.}~\bibnamefont {Jagielski}}, \ and\ \bibinfo {author} {\bibfnamefont {M.}~\bibnamefont {Alava}},\ }\href {\doibase https://doi.org/10.1016/j.matdes.2022.110639} {\bibfield  {journal} {\bibinfo  {journal} {Materials \& Design}\ }\textbf {\bibinfo
  {volume} {217}},\ \bibinfo {pages} {110639} (\bibinfo {year} {2022})}\BibitemShut {NoStop}%
\bibitem [{\citenamefont {Frydrych}\ \emph {et~al.}(2023)\citenamefont {Frydrych}, \citenamefont {Dominguez-Gutierrez}, \citenamefont {Alava},\ and\ \citenamefont {Papanikolaou}}]{FRYDRYCH2023104644}%
  \BibitemOpen
  \bibfield  {author} {\bibinfo {author} {\bibfnamefont {K.}~\bibnamefont {Frydrych}}, \bibinfo {author} {\bibfnamefont {F.}~\bibnamefont {Dominguez-Gutierrez}}, \bibinfo {author} {\bibfnamefont {M.}~\bibnamefont {Alava}}, \ and\ \bibinfo {author} {\bibfnamefont {S.}~\bibnamefont {Papanikolaou}},\ }\href {\doibase https://doi.org/10.1016/j.mechmat.2023.104644} {\bibfield  {journal} {\bibinfo  {journal} {Mechanics of Materials}\ }\textbf {\bibinfo {volume} {181}},\ \bibinfo {pages} {104644} (\bibinfo {year} {2023})}\BibitemShut {NoStop}%
\bibitem [{\citenamefont {Stasiak}\ \emph {et~al.}(2022)\citenamefont {Stasiak}, \citenamefont {Oleszak},\ and\ \citenamefont {Fraczkiewicz}}]{stasiak2022effects}%
  \BibitemOpen
  \bibfield  {author} {\bibinfo {author} {\bibfnamefont {T.}~\bibnamefont {Stasiak}}, \bibinfo {author} {\bibfnamefont {D.}~\bibnamefont {Oleszak}}, \ and\ \bibinfo {author} {\bibfnamefont {A.}~\bibnamefont {Fraczkiewicz}},\ }\href {\doibase 10.1007/s11837-022-05543-2} {\bibfield  {journal} {\bibinfo  {journal} {JOM}\ }\textbf {\bibinfo {volume} {74}},\ \bibinfo {pages} {4842} (\bibinfo {year} {2022})}\BibitemShut {NoStop}%
\bibitem [{\citenamefont {Lu}\ \emph {et~al.}(2016)\citenamefont {Lu}, \citenamefont {Niu}, \citenamefont {Chen} \emph {et~al.}}]{lu2016enhancing}%
  \BibitemOpen
  \bibfield  {author} {\bibinfo {author} {\bibfnamefont {C.}~\bibnamefont {Lu}}, \bibinfo {author} {\bibfnamefont {L.}~\bibnamefont {Niu}}, \bibinfo {author} {\bibfnamefont {N.}~\bibnamefont {Chen}},  \emph {et~al.},\ }\href {\doibase 10.1038/ncomms13564} {\bibfield  {journal} {\bibinfo  {journal} {Nature Communications}\ }\textbf {\bibinfo {volume} {7}},\ \bibinfo {pages} {13564} (\bibinfo {year} {2016})}\BibitemShut {NoStop}%
\bibitem [{\citenamefont {Jin}\ \emph {et~al.}(2016)\citenamefont {Jin}, \citenamefont {Lu}, \citenamefont {Wang}, \citenamefont {Qu}, \citenamefont {Weber}, \citenamefont {Zhang},\ and\ \citenamefont {Bei}}]{JIN201665}%
  \BibitemOpen
  \bibfield  {author} {\bibinfo {author} {\bibfnamefont {K.}~\bibnamefont {Jin}}, \bibinfo {author} {\bibfnamefont {C.}~\bibnamefont {Lu}}, \bibinfo {author} {\bibfnamefont {L.}~\bibnamefont {Wang}}, \bibinfo {author} {\bibfnamefont {J.}~\bibnamefont {Qu}}, \bibinfo {author} {\bibfnamefont {W.}~\bibnamefont {Weber}}, \bibinfo {author} {\bibfnamefont {Y.}~\bibnamefont {Zhang}}, \ and\ \bibinfo {author} {\bibfnamefont {H.}~\bibnamefont {Bei}},\ }\href {\doibase https://doi.org/10.1016/j.scriptamat.2016.03.030} {\bibfield  {journal} {\bibinfo  {journal} {Scripta Materialia}\ }\textbf {\bibinfo {volume} {119}},\ \bibinfo {pages} {65} (\bibinfo {year} {2016})}\BibitemShut {NoStop}%
\bibitem [{\citenamefont {Daramola}\ \emph {et~al.}(2022)\citenamefont {Daramola}, \citenamefont {Bonny}, \citenamefont {Adjanor}, \citenamefont {Domain}, \citenamefont {Monnet},\ and\ \citenamefont {Fraczkiewicz}}]{DARAMOLA2022111165}%
  \BibitemOpen
  \bibfield  {author} {\bibinfo {author} {\bibfnamefont {A.}~\bibnamefont {Daramola}}, \bibinfo {author} {\bibfnamefont {G.}~\bibnamefont {Bonny}}, \bibinfo {author} {\bibfnamefont {G.}~\bibnamefont {Adjanor}}, \bibinfo {author} {\bibfnamefont {C.}~\bibnamefont {Domain}}, \bibinfo {author} {\bibfnamefont {G.}~\bibnamefont {Monnet}}, \ and\ \bibinfo {author} {\bibfnamefont {A.}~\bibnamefont {Fraczkiewicz}},\ }\href {\doibase https://doi.org/10.1016/j.commatsci.2021.111165} {\bibfield  {journal} {\bibinfo  {journal} {Computational Materials Science}\ }\textbf {\bibinfo {volume} {203}},\ \bibinfo {pages} {111165} (\bibinfo {year} {2022})}\BibitemShut {NoStop}%
\bibitem [{\citenamefont {Zhong}\ \emph {et~al.}(2023)\citenamefont {Zhong}, \citenamefont {Hayakawa}, \citenamefont {Xu}, \citenamefont {An}, \citenamefont {Borisevich}, \citenamefont {Cicotte}, \citenamefont {George},\ and\ \citenamefont {Yang}}]{ZHONG2023103663}%
  \BibitemOpen
  \bibfield  {author} {\bibinfo {author} {\bibfnamefont {W.}~\bibnamefont {Zhong}}, \bibinfo {author} {\bibfnamefont {S.}~\bibnamefont {Hayakawa}}, \bibinfo {author} {\bibfnamefont {H.}~\bibnamefont {Xu}}, \bibinfo {author} {\bibfnamefont {K.}~\bibnamefont {An}}, \bibinfo {author} {\bibfnamefont {A.~Y.}\ \bibnamefont {Borisevich}}, \bibinfo {author} {\bibfnamefont {J.~L.}\ \bibnamefont {Cicotte}}, \bibinfo {author} {\bibfnamefont {E.~P.}\ \bibnamefont {George}}, \ and\ \bibinfo {author} {\bibfnamefont {Y.}~\bibnamefont {Yang}},\ }\href {\doibase https://doi.org/10.1016/j.ijplas.2023.103663} {\bibfield  {journal} {\bibinfo  {journal} {International Journal of Plasticity}\ }\textbf {\bibinfo {volume} {167}},\ \bibinfo {pages} {103663} (\bibinfo {year} {2023})}\BibitemShut {NoStop}%
\bibitem [{\citenamefont {Schuh}(2006)}]{SCHUH200632}%
  \BibitemOpen
  \bibfield  {author} {\bibinfo {author} {\bibfnamefont {C.~A.}\ \bibnamefont {Schuh}},\ }\href {\doibase https://doi.org/10.1016/S1369-7021(06)71495-X} {\bibfield  {journal} {\bibinfo  {journal} {Materials Today}\ }\textbf {\bibinfo {volume} {9}},\ \bibinfo {pages} {32} (\bibinfo {year} {2006})}\BibitemShut {NoStop}%
\bibitem [{\citenamefont {Varillas}\ \emph {et~al.}(2017)\citenamefont {Varillas}, \citenamefont {Ocenasek}, \citenamefont {Torner},\ and\ \citenamefont {Alcala}}]{VARILLAS2017431}%
  \BibitemOpen
  \bibfield  {author} {\bibinfo {author} {\bibfnamefont {J.}~\bibnamefont {Varillas}}, \bibinfo {author} {\bibfnamefont {J.}~\bibnamefont {Ocenasek}}, \bibinfo {author} {\bibfnamefont {J.}~\bibnamefont {Torner}}, \ and\ \bibinfo {author} {\bibfnamefont {J.}~\bibnamefont {Alcala}},\ }\href {\doibase https://doi.org/10.1016/j.actamat.2016.11.067} {\bibfield  {journal} {\bibinfo  {journal} {Acta Materialia}\ }\textbf {\bibinfo {volume} {125}},\ \bibinfo {pages} {431} (\bibinfo {year} {2017})}\BibitemShut {NoStop}%
\bibitem [{\citenamefont {Pathak}\ and\ \citenamefont {Kalidindi}(2015)}]{PATHAK20151}%
  \BibitemOpen
  \bibfield  {author} {\bibinfo {author} {\bibfnamefont {S.}~\bibnamefont {Pathak}}\ and\ \bibinfo {author} {\bibfnamefont {S.~R.}\ \bibnamefont {Kalidindi}},\ }\href@noop {} {\bibfield  {journal} {\bibinfo  {journal} {Materials Science and Engineering: R: Reports}\ }\textbf {\bibinfo {volume} {91}},\ \bibinfo {pages} {1} (\bibinfo {year} {2015})}\BibitemShut {NoStop}%
\bibitem [{\citenamefont {Yang}\ \emph {et~al.}(2022)\citenamefont {Yang}, \citenamefont {Chen}, \citenamefont {Miller}, \citenamefont {Weber}, \citenamefont {Bei},\ and\ \citenamefont {Zhang}}]{YANG2022143685}%
  \BibitemOpen
  \bibfield  {author} {\bibinfo {author} {\bibfnamefont {L.}~\bibnamefont {Yang}}, \bibinfo {author} {\bibfnamefont {Y.}~\bibnamefont {Chen}}, \bibinfo {author} {\bibfnamefont {J.}~\bibnamefont {Miller}}, \bibinfo {author} {\bibfnamefont {W.~J.}\ \bibnamefont {Weber}}, \bibinfo {author} {\bibfnamefont {H.}~\bibnamefont {Bei}}, \ and\ \bibinfo {author} {\bibfnamefont {Y.}~\bibnamefont {Zhang}},\ }\href {\doibase https://doi.org/10.1016/j.msea.2022.143685} {\bibfield  {journal} {\bibinfo  {journal} {Materials Science and Engineering: A}\ }\textbf {\bibinfo {volume} {856}},\ \bibinfo {pages} {143685} (\bibinfo {year} {2022})}\BibitemShut {NoStop}%
\bibitem [{\citenamefont {Pathak}\ \emph {et~al.}(2016)\citenamefont {Pathak}, \citenamefont {Kalidindi},\ and\ \citenamefont {Mara}}]{PATHAK2016241}%
  \BibitemOpen
  \bibfield  {author} {\bibinfo {author} {\bibfnamefont {S.}~\bibnamefont {Pathak}}, \bibinfo {author} {\bibfnamefont {S.~R.}\ \bibnamefont {Kalidindi}}, \ and\ \bibinfo {author} {\bibfnamefont {N.~A.}\ \bibnamefont {Mara}},\ }\href {\doibase https://doi.org/10.1016/j.scriptamat.2015.10.035} {\bibfield  {journal} {\bibinfo  {journal} {Scripta Materialia}\ }\textbf {\bibinfo {volume} {113}},\ \bibinfo {pages} {241} (\bibinfo {year} {2016})}\BibitemShut {NoStop}%
\bibitem [{\citenamefont {Remington}\ \emph {et~al.}(2014)\citenamefont {Remington}, \citenamefont {Ruestes}, \citenamefont {Bringa}, \citenamefont {Remington}, \citenamefont {Lu}, \citenamefont {Kad},\ and\ \citenamefont {Meyers}}]{REMINGTON2014378}%
  \BibitemOpen
  \bibfield  {author} {\bibinfo {author} {\bibfnamefont {T.}~\bibnamefont {Remington}}, \bibinfo {author} {\bibfnamefont {C.}~\bibnamefont {Ruestes}}, \bibinfo {author} {\bibfnamefont {E.}~\bibnamefont {Bringa}}, \bibinfo {author} {\bibfnamefont {B.}~\bibnamefont {Remington}}, \bibinfo {author} {\bibfnamefont {C.}~\bibnamefont {Lu}}, \bibinfo {author} {\bibfnamefont {B.}~\bibnamefont {Kad}}, \ and\ \bibinfo {author} {\bibfnamefont {M.}~\bibnamefont {Meyers}},\ }\href {\doibase https://doi.org/10.1016/j.actamat.2014.06.058} {\bibfield  {journal} {\bibinfo  {journal} {Acta Materialia}\ }\textbf {\bibinfo {volume} {78}},\ \bibinfo {pages} {378} (\bibinfo {year} {2014})}\BibitemShut {NoStop}%
\bibitem [{\citenamefont {Wyszkowska}\ \emph {et~al.}(2023)\citenamefont {Wyszkowska}, \citenamefont {Mieszczynski}, \citenamefont {Kurpaska}, \citenamefont {Azarov}, \citenamefont {Jalswik}, \citenamefont {Kosinska}, \citenamefont {Chrominski}, \citenamefont {Diduszko}, \citenamefont {Huo}, \citenamefont {Cienlik},\ and\ \citenamefont {Jagielski}}]{D2NR06178C}%
  \BibitemOpen
  \bibfield  {author} {\bibinfo {author} {\bibfnamefont {E.}~\bibnamefont {Wyszkowska}}, \bibinfo {author} {\bibfnamefont {C.}~\bibnamefont {Mieszczynski}}, \bibinfo {author} {\bibfnamefont {L.}~\bibnamefont {Kurpaska}}, \bibinfo {author} {\bibfnamefont {A.}~\bibnamefont {Azarov}}, \bibinfo {author} {\bibfnamefont {I.}~\bibnamefont {Jalswik}}, \bibinfo {author} {\bibfnamefont {A.}~\bibnamefont {Kosinska}}, \bibinfo {author} {\bibfnamefont {W.}~\bibnamefont {Chrominski}}, \bibinfo {author} {\bibfnamefont {R.}~\bibnamefont {Diduszko}}, \bibinfo {author} {\bibfnamefont {W.~Y.}\ \bibnamefont {Huo}}, \bibinfo {author} {\bibfnamefont {I.}~\bibnamefont {Cienlik}}, \ and\ \bibinfo {author} {\bibfnamefont {J.}~\bibnamefont {Jagielski}},\ }\href {\doibase 10.1039/D2NR06178C} {\bibfield  {journal} {\bibinfo  {journal} {Nanoscale}\ }\textbf {\bibinfo {volume} {15}},\ \bibinfo {pages} {4870} (\bibinfo {year} {2023})}\BibitemShut {NoStop}%
\bibitem [{\citenamefont {Li}\ \emph {et~al.}(2020)\citenamefont {Li}, \citenamefont {Gao}, \citenamefont {Brand}, \citenamefont {Hiller},\ and\ \citenamefont {Wolff}}]{Li_2020}%
  \BibitemOpen
  \bibfield  {author} {\bibinfo {author} {\bibfnamefont {Z.}~\bibnamefont {Li}}, \bibinfo {author} {\bibfnamefont {S.}~\bibnamefont {Gao}}, \bibinfo {author} {\bibfnamefont {U.}~\bibnamefont {Brand}}, \bibinfo {author} {\bibfnamefont {K.}~\bibnamefont {Hiller}}, \ and\ \bibinfo {author} {\bibfnamefont {H.}~\bibnamefont {Wolff}},\ }\href {\doibase 10.1088/1361-6528/ab88ed} {\bibfield  {journal} {\bibinfo  {journal} {Nanotechnology}\ }\textbf {\bibinfo {volume} {31}},\ \bibinfo {pages} {305502} (\bibinfo {year} {2020})}\BibitemShut {NoStop}%
\bibitem [{\citenamefont {Dominguez-Gutierrez}(2022)}]{DOMINGUEZGUTIERREZ202238}%
  \BibitemOpen
  \bibfield  {author} {\bibinfo {author} {\bibfnamefont {F.}~\bibnamefont {Dominguez-Gutierrez}},\ }\href@noop {} {\bibfield  {journal} {\bibinfo  {journal} {Nuclear Instruments and Methods in Physics Research Section B: Beam Interactions with Materials and Atoms}\ }\textbf {\bibinfo {volume} {512}},\ \bibinfo {pages} {38} (\bibinfo {year} {2022})}\BibitemShut {NoStop}%
\bibitem [{\citenamefont {Hu}\ \emph {et~al.}(2022)\citenamefont {Hu}, \citenamefont {Fu}, \citenamefont {Liang}, \citenamefont {Weng}, \citenamefont {Chen}, \citenamefont {Zhao},\ and\ \citenamefont {Peng}}]{hu2022formation}%
  \BibitemOpen
  \bibfield  {author} {\bibinfo {author} {\bibfnamefont {S.}~\bibnamefont {Hu}}, \bibinfo {author} {\bibfnamefont {T.}~\bibnamefont {Fu}}, \bibinfo {author} {\bibfnamefont {Q.}~\bibnamefont {Liang}}, \bibinfo {author} {\bibfnamefont {S.}~\bibnamefont {Weng}}, \bibinfo {author} {\bibfnamefont {X.}~\bibnamefont {Chen}}, \bibinfo {author} {\bibfnamefont {Y.}~\bibnamefont {Zhao}}, \ and\ \bibinfo {author} {\bibfnamefont {X.}~\bibnamefont {Peng}},\ }\href@noop {} {\bibfield  {journal} {\bibinfo  {journal} {Frontiers in Materials}\ }\textbf {\bibinfo {volume} {8}},\ \bibinfo {pages} {813382} (\bibinfo {year} {2022})}\BibitemShut {NoStop}%
\bibitem [{\citenamefont {Chandan}\ \emph {et~al.}(2021)\citenamefont {Chandan}, \citenamefont {Tripathy}, \citenamefont {Sen}, \citenamefont {Ghosh},\ and\ \citenamefont {Chowdhury}}]{chandan2021temperature}%
  \BibitemOpen
  \bibfield  {author} {\bibinfo {author} {\bibfnamefont {A.}~\bibnamefont {Chandan}}, \bibinfo {author} {\bibfnamefont {S.}~\bibnamefont {Tripathy}}, \bibinfo {author} {\bibfnamefont {B.}~\bibnamefont {Sen}}, \bibinfo {author} {\bibfnamefont {M.}~\bibnamefont {Ghosh}}, \ and\ \bibinfo {author} {\bibfnamefont {S.~G.}\ \bibnamefont {Chowdhury}},\ }\href@noop {} {\bibfield  {journal} {\bibinfo  {journal} {Scripta Materialia}\ }\textbf {\bibinfo {volume} {199}},\ \bibinfo {pages} {113891} (\bibinfo {year} {2021})}\BibitemShut {NoStop}%
\bibitem [{\citenamefont {Huang}\ \emph {et~al.}(2015)\citenamefont {Huang}, \citenamefont {Li}, \citenamefont {Lu}, \citenamefont {Tian}, \citenamefont {Shen}, \citenamefont {Holmstr{\"o}m},\ and\ \citenamefont {Vitos}}]{huang2015temperature}%
  \BibitemOpen
  \bibfield  {author} {\bibinfo {author} {\bibfnamefont {S.}~\bibnamefont {Huang}}, \bibinfo {author} {\bibfnamefont {W.}~\bibnamefont {Li}}, \bibinfo {author} {\bibfnamefont {S.}~\bibnamefont {Lu}}, \bibinfo {author} {\bibfnamefont {F.}~\bibnamefont {Tian}}, \bibinfo {author} {\bibfnamefont {J.}~\bibnamefont {Shen}}, \bibinfo {author} {\bibfnamefont {E.}~\bibnamefont {Holmstr{\"o}m}}, \ and\ \bibinfo {author} {\bibfnamefont {L.}~\bibnamefont {Vitos}},\ }\href@noop {} {\bibfield  {journal} {\bibinfo  {journal} {Scripta Materialia}\ }\textbf {\bibinfo {volume} {108}},\ \bibinfo {pages} {44} (\bibinfo {year} {2015})}\BibitemShut {NoStop}%
\bibitem [{\citenamefont {Thompson}\ \emph {et~al.}(2022)\citenamefont {Thompson}, \citenamefont {Aktulga}, \citenamefont {Berger}, \citenamefont {Bolintineanu}, \citenamefont {Brown}, \citenamefont {Crozier}, \citenamefont {{in 't Veld}}, \citenamefont {Kohlmeyer}, \citenamefont {Moore}, \citenamefont {Nguyen}, \citenamefont {Shan}, \citenamefont {Stevens}, \citenamefont {Tranchida}, \citenamefont {Trott},\ and\ \citenamefont {Plimpton}}]{THOMPSON2022108171}%
  \BibitemOpen
  \bibfield  {author} {\bibinfo {author} {\bibfnamefont {A.~P.}\ \bibnamefont {Thompson}}, \bibinfo {author} {\bibfnamefont {H.~M.}\ \bibnamefont {Aktulga}}, \bibinfo {author} {\bibfnamefont {R.}~\bibnamefont {Berger}}, \bibinfo {author} {\bibfnamefont {D.~S.}\ \bibnamefont {Bolintineanu}}, \bibinfo {author} {\bibfnamefont {W.~M.}\ \bibnamefont {Brown}}, \bibinfo {author} {\bibfnamefont {P.~S.}\ \bibnamefont {Crozier}}, \bibinfo {author} {\bibfnamefont {P.~J.}\ \bibnamefont {{in 't Veld}}}, \bibinfo {author} {\bibfnamefont {A.}~\bibnamefont {Kohlmeyer}}, \bibinfo {author} {\bibfnamefont {S.~G.}\ \bibnamefont {Moore}}, \bibinfo {author} {\bibfnamefont {T.~D.}\ \bibnamefont {Nguyen}}, \bibinfo {author} {\bibfnamefont {R.}~\bibnamefont {Shan}}, \bibinfo {author} {\bibfnamefont {M.~J.}\ \bibnamefont {Stevens}}, \bibinfo {author} {\bibfnamefont {J.}~\bibnamefont {Tranchida}}, \bibinfo {author} {\bibfnamefont {C.}~\bibnamefont {Trott}}, \ and\ \bibinfo {author} {\bibfnamefont {S.~J.}\ \bibnamefont {Plimpton}},\
  }\href {\doibase https://doi.org/10.1016/j.cpc.2021.108171} {\bibfield  {journal} {\bibinfo  {journal} {Computer Physics Communications}\ }\textbf {\bibinfo {volume} {271}},\ \bibinfo {pages} {108171} (\bibinfo {year} {2022})}\BibitemShut {NoStop}%
\bibitem [{\citenamefont {Choi}\ \emph {et~al.}(2018)\citenamefont {Choi}, \citenamefont {Jo}, \citenamefont {Sohn} \emph {et~al.}}]{Choietal}%
  \BibitemOpen
  \bibfield  {author} {\bibinfo {author} {\bibfnamefont {W.~M.}\ \bibnamefont {Choi}}, \bibinfo {author} {\bibfnamefont {Y.}~\bibnamefont {Jo}}, \bibinfo {author} {\bibfnamefont {S.}~\bibnamefont {Sohn}},  \emph {et~al.},\ }\href@noop {} {\bibfield  {journal} {\bibinfo  {journal} {npj Comput Mater}\ }\textbf {\bibinfo {volume} {4}},\ \bibinfo {pages} {1} (\bibinfo {year} {2018})}\BibitemShut {NoStop}%
\bibitem [{\citenamefont {Naghdi}\ \emph {et~al.}(2022)\citenamefont {Naghdi}, \citenamefont {Dominguez-Gutierrez}, \citenamefont {Huo}, \citenamefont {Karimi},\ and\ \citenamefont {Papanikolaou}}]{naghdi2022dynamic}%
  \BibitemOpen
  \bibfield  {author} {\bibinfo {author} {\bibfnamefont {A.}~\bibnamefont {Naghdi}}, \bibinfo {author} {\bibfnamefont {F.~J.}\ \bibnamefont {Dominguez-Gutierrez}}, \bibinfo {author} {\bibfnamefont {W.~Y.}\ \bibnamefont {Huo}}, \bibinfo {author} {\bibfnamefont {K.}~\bibnamefont {Karimi}}, \ and\ \bibinfo {author} {\bibfnamefont {S.}~\bibnamefont {Papanikolaou}},\ }\href@noop {} {\enquote {\bibinfo {title} {Dynamic nanoindentation and short-range order in equiatomic nicocr medium entropy alloy lead to novel density wave ordering},}\ } (\bibinfo {year} {2022}),\ \Eprint {http://arxiv.org/abs/2211.05436} {arXiv:2211.05436 [cond-mat.mtrl-sci]} \BibitemShut {NoStop}%
\bibitem [{\citenamefont {Gu\'enol\'e}\ \emph {et~al.}(2020)\citenamefont {Gu\'enol\'e}, \citenamefont {N\"ohring}, \citenamefont {Vaid}, \citenamefont {Houll\'e}, \citenamefont {Xie}, \citenamefont {Prakash},\ and\ \citenamefont {Bitzek}}]{GUENOLE2020109584}%
  \BibitemOpen
  \bibfield  {author} {\bibinfo {author} {\bibfnamefont {J.}~\bibnamefont {Gu\'enol\'e}}, \bibinfo {author} {\bibfnamefont {W.~G.}\ \bibnamefont {N\"ohring}}, \bibinfo {author} {\bibfnamefont {A.}~\bibnamefont {Vaid}}, \bibinfo {author} {\bibfnamefont {F.}~\bibnamefont {Houll\'e}}, \bibinfo {author} {\bibfnamefont {Z.}~\bibnamefont {Xie}}, \bibinfo {author} {\bibfnamefont {A.}~\bibnamefont {Prakash}}, \ and\ \bibinfo {author} {\bibfnamefont {E.}~\bibnamefont {Bitzek}},\ }\href {\doibase https://doi.org/10.1016/j.commatsci.2020.109584} {\bibfield  {journal} {\bibinfo  {journal} {Computational Materials Science}\ }\textbf {\bibinfo {volume} {175}},\ \bibinfo {pages} {109584} (\bibinfo {year} {2020})}\BibitemShut {NoStop}%
\bibitem [{\citenamefont {Dominguez-Gutierrez}\ \emph {et~al.}(2021)\citenamefont {Dominguez-Gutierrez}, \citenamefont {Papanikolaou}, \citenamefont {Esfandiarpour}, \citenamefont {Sobkowicz},\ and\ \citenamefont {Alava}}]{DOMINGUEZGUTIERREZ2021141912}%
  \BibitemOpen
  \bibfield  {author} {\bibinfo {author} {\bibfnamefont {F.}~\bibnamefont {Dominguez-Gutierrez}}, \bibinfo {author} {\bibfnamefont {S.}~\bibnamefont {Papanikolaou}}, \bibinfo {author} {\bibfnamefont {A.}~\bibnamefont {Esfandiarpour}}, \bibinfo {author} {\bibfnamefont {P.}~\bibnamefont {Sobkowicz}}, \ and\ \bibinfo {author} {\bibfnamefont {M.}~\bibnamefont {Alava}},\ }\href {\doibase https://doi.org/10.1016/j.msea.2021.141912} {\bibfield  {journal} {\bibinfo  {journal} {Materials Science and Engineering: A}\ }\textbf {\bibinfo {volume} {826}},\ \bibinfo {pages} {141912} (\bibinfo {year} {2021})}\BibitemShut {NoStop}%
\bibitem [{\citenamefont {Xu}\ \emph {et~al.}(2024)\citenamefont {Xu}, \citenamefont {Zaborowska}, \citenamefont {Mulewska}, \citenamefont {Huo}, \citenamefont {Karimi}, \citenamefont {Dominguez-Gutierrez}, \citenamefont {Kurpaska}, \citenamefont {Alava},\ and\ \citenamefont {Papanikolaou}}]{XU2024112733}%
  \BibitemOpen
  \bibfield  {author} {\bibinfo {author} {\bibfnamefont {Q.}~\bibnamefont {Xu}}, \bibinfo {author} {\bibfnamefont {A.}~\bibnamefont {Zaborowska}}, \bibinfo {author} {\bibfnamefont {K.}~\bibnamefont {Mulewska}}, \bibinfo {author} {\bibfnamefont {W.}~\bibnamefont {Huo}}, \bibinfo {author} {\bibfnamefont {K.}~\bibnamefont {Karimi}}, \bibinfo {author} {\bibfnamefont {F.~J.}\ \bibnamefont {Dominguez-Gutierrez}}, \bibinfo {author} {\bibfnamefont {L.}~\bibnamefont {Kurpaska}}, \bibinfo {author} {\bibfnamefont {M.~J.}\ \bibnamefont {Alava}}, \ and\ \bibinfo {author} {\bibfnamefont {S.}~\bibnamefont {Papanikolaou}},\ }\href {\doibase https://doi.org/10.1016/j.vacuum.2023.112733} {\bibfield  {journal} {\bibinfo  {journal} {Vacuum}\ }\textbf {\bibinfo {volume} {219}},\ \bibinfo {pages} {112733} (\bibinfo {year} {2024})}\BibitemShut {NoStop}%
\bibitem [{\citenamefont {Silcox}\ and\ \citenamefont {Hirsch}(1959)}]{silcox1959direct}%
  \BibitemOpen
  \bibfield  {author} {\bibinfo {author} {\bibfnamefont {J.}~\bibnamefont {Silcox}}\ and\ \bibinfo {author} {\bibfnamefont {P.}~\bibnamefont {Hirsch}},\ }\href {\doibase https://doi.org/10.1080/14786435908238228} {\bibfield  {journal} {\bibinfo  {journal} {Philosophical Magazine}\ }\textbf {\bibinfo {volume} {4}},\ \bibinfo {pages} {72} (\bibinfo {year} {1959})}\BibitemShut {NoStop}%
\bibitem [{\citenamefont {Landeiro Dos~Reis}\ \emph {et~al.}(2020)\citenamefont {Landeiro Dos~Reis}, \citenamefont {Proville}, \citenamefont {Marinica},\ and\ \citenamefont {Sauzay}}]{PhysRevMaterials.4.103603}%
  \BibitemOpen
  \bibfield  {author} {\bibinfo {author} {\bibfnamefont {M.}~\bibnamefont {Landeiro Dos~Reis}}, \bibinfo {author} {\bibfnamefont {L.}~\bibnamefont {Proville}}, \bibinfo {author} {\bibfnamefont {M.-C.}\ \bibnamefont {Marinica}}, \ and\ \bibinfo {author} {\bibfnamefont {M.}~\bibnamefont {Sauzay}},\ }\href {\doibase 10.1103/PhysRevMaterials.4.103603} {\bibfield  {journal} {\bibinfo  {journal} {Phys. Rev. Mater.}\ }\textbf {\bibinfo {volume} {4}},\ \bibinfo {pages} {103603} (\bibinfo {year} {2020})}\BibitemShut {NoStop}%
\bibitem [{\citenamefont {Dom\'{\i}nguez-Guti\'errez}\ \emph {et~al.}(2023)\citenamefont {Dom\'{\i}nguez-Guti\'errez}, \citenamefont {Grigorev}, \citenamefont {Naghdi}, \citenamefont {Byggm\"astar}, \citenamefont {Wei}, \citenamefont {Swinburne}, \citenamefont {Papanikolaou},\ and\ \citenamefont {Alava}}]{PhysRevMaterials.7.043603}%
  \BibitemOpen
  \bibfield  {author} {\bibinfo {author} {\bibfnamefont {F.~J.}\ \bibnamefont {Dom\'{\i}nguez-Guti\'errez}}, \bibinfo {author} {\bibfnamefont {P.}~\bibnamefont {Grigorev}}, \bibinfo {author} {\bibfnamefont {A.}~\bibnamefont {Naghdi}}, \bibinfo {author} {\bibfnamefont {J.}~\bibnamefont {Byggm\"astar}}, \bibinfo {author} {\bibfnamefont {G.~Y.}\ \bibnamefont {Wei}}, \bibinfo {author} {\bibfnamefont {T.~D.}\ \bibnamefont {Swinburne}}, \bibinfo {author} {\bibfnamefont {S.}~\bibnamefont {Papanikolaou}}, \ and\ \bibinfo {author} {\bibfnamefont {M.~J.}\ \bibnamefont {Alava}},\ }\href {\doibase 10.1103/PhysRevMaterials.7.043603} {\bibfield  {journal} {\bibinfo  {journal} {Phys. Rev. Mater.}\ }\textbf {\bibinfo {volume} {7}},\ \bibinfo {pages} {043603} (\bibinfo {year} {2023})}\BibitemShut {NoStop}%
\bibitem [{\citenamefont {Varillas-Delgado}\ and\ \citenamefont {Alcala~Cabrelles}(2019)}]{JavVarilla}%
  \BibitemOpen
  \bibfield  {author} {\bibinfo {author} {\bibfnamefont {J.}~\bibnamefont {Varillas-Delgado}}\ and\ \bibinfo {author} {\bibfnamefont {J.}~\bibnamefont {Alcala~Cabrelles}},\ }\emph {\bibinfo {title} {A molecular dynamics study of nanocontact plasticity and dislocation avalanches in FCC and BCC crystals}},\ \href {https://www.researchgate.net/publication/336876966_A_Molecular_Dynamics_Study_of_Nanocontact_Plasticity_and_Dislocation_Avalanches_in_FCC_and_BCC_Crystals} {\bibinfo {type} {{PhD} dissertation}},\ \bibinfo  {school} {Universitat Politècnica de Catalunya. Departament de Ciència dels Materials i Enginyeria Metal·lúrgica} (\bibinfo {year} {2019})\BibitemShut {NoStop}%
\bibitem [{\citenamefont {Stukowski}(2010)}]{ovito}%
  \BibitemOpen
  \bibfield  {author} {\bibinfo {author} {\bibfnamefont {A.}~\bibnamefont {Stukowski}},\ }\href {\doibase {10.1088/0965-0393/18/1/015012}} {\bibfield  {journal} {\bibinfo  {journal} {{Modelling and simulation in materials science and engineering}}\ }\textbf {\bibinfo {volume} {{18}}} (\bibinfo {year} {{2010}}),\ {10.1088/0965-0393/18/1/015012}}\BibitemShut {NoStop}%
\bibitem [{\citenamefont {Stukowski}\ \emph {et~al.}(2012)\citenamefont {Stukowski}, \citenamefont {Bulatov},\ and\ \citenamefont {Arsenlis}}]{Stukowski_2012}%
  \BibitemOpen
  \bibfield  {author} {\bibinfo {author} {\bibfnamefont {A.}~\bibnamefont {Stukowski}}, \bibinfo {author} {\bibfnamefont {V.~V.}\ \bibnamefont {Bulatov}}, \ and\ \bibinfo {author} {\bibfnamefont {A.}~\bibnamefont {Arsenlis}},\ }\href {\doibase 10.1088/0965-0393/20/8/085007} {\bibfield  {journal} {\bibinfo  {journal} {Modelling and Simulation in Materials Science and Engineering}\ }\textbf {\bibinfo {volume} {20}},\ \bibinfo {pages} {085007} (\bibinfo {year} {2012})}\BibitemShut {NoStop}%
\bibitem [{\citenamefont {Daphalapurkar}\ and\ \citenamefont {Ramesh}(2012)}]{DAPHALAPURKAR2012277}%
  \BibitemOpen
  \bibfield  {author} {\bibinfo {author} {\bibfnamefont {N.~P.}\ \bibnamefont {Daphalapurkar}}\ and\ \bibinfo {author} {\bibfnamefont {K.}~\bibnamefont {Ramesh}},\ }\href {\doibase https://doi.org/10.1016/j.jmps.2011.10.009} {\bibfield  {journal} {\bibinfo  {journal} {Journal of the Mechanics and Physics of Solids}\ }\textbf {\bibinfo {volume} {60}},\ \bibinfo {pages} {277} (\bibinfo {year} {2012})}\BibitemShut {NoStop}%
\bibitem [{\citenamefont {Kalidindi}(1998)}]{KALIDINDI1998267}%
  \BibitemOpen
  \bibfield  {author} {\bibinfo {author} {\bibfnamefont {S.~R.}\ \bibnamefont {Kalidindi}},\ }\href {\doibase https://doi.org/10.1016/S0022-5096(97)00051-3} {\bibfield  {journal} {\bibinfo  {journal} {Journal of the Mechanics and Physics of Solids}\ }\textbf {\bibinfo {volume} {46}},\ \bibinfo {pages} {267} (\bibinfo {year} {1998})}\BibitemShut {NoStop}%
\bibitem [{\citenamefont {Shen}\ and\ \citenamefont {Spearot}(2021)}]{shen2021mobility}%
  \BibitemOpen
  \bibfield  {author} {\bibinfo {author} {\bibfnamefont {Y.}~\bibnamefont {Shen}}\ and\ \bibinfo {author} {\bibfnamefont {D.~E.}\ \bibnamefont {Spearot}},\ }\href@noop {} {\bibfield  {journal} {\bibinfo  {journal} {Modelling and Simulation in Materials Science and Engineering}\ }\textbf {\bibinfo {volume} {29}},\ \bibinfo {pages} {085017} (\bibinfo {year} {2021})}\BibitemShut {NoStop}%
\bibitem [{\citenamefont {Alhafez}\ \emph {et~al.}(2019)\citenamefont {Alhafez}, \citenamefont {Ruestes}, \citenamefont {Bringa},\ and\ \citenamefont {Urbassek}}]{alhafez2019nanoindentation}%
  \BibitemOpen
  \bibfield  {author} {\bibinfo {author} {\bibfnamefont {I.~A.}\ \bibnamefont {Alhafez}}, \bibinfo {author} {\bibfnamefont {C.~J.}\ \bibnamefont {Ruestes}}, \bibinfo {author} {\bibfnamefont {E.~M.}\ \bibnamefont {Bringa}}, \ and\ \bibinfo {author} {\bibfnamefont {H.~M.}\ \bibnamefont {Urbassek}},\ }\href@noop {} {\bibfield  {journal} {\bibinfo  {journal} {Journal of Alloys and Compounds}\ }\textbf {\bibinfo {volume} {803}},\ \bibinfo {pages} {618} (\bibinfo {year} {2019})}\BibitemShut {NoStop}%
\bibitem [{\citenamefont {Dominguez-Gutierrez}\ \emph {et~al.}(2022)\citenamefont {Dominguez-Gutierrez}, \citenamefont {Ustrzycka}, \citenamefont {Xu}, \citenamefont {Alvarez-Donado}, \citenamefont {Papanikolaou},\ and\ \citenamefont {Alava}}]{Dominguez-Gutierrez_2022}%
  \BibitemOpen
  \bibfield  {author} {\bibinfo {author} {\bibfnamefont {F.~J.}\ \bibnamefont {Dominguez-Gutierrez}}, \bibinfo {author} {\bibfnamefont {A.}~\bibnamefont {Ustrzycka}}, \bibinfo {author} {\bibfnamefont {Q.~Q.}\ \bibnamefont {Xu}}, \bibinfo {author} {\bibfnamefont {R.}~\bibnamefont {Alvarez-Donado}}, \bibinfo {author} {\bibfnamefont {S.}~\bibnamefont {Papanikolaou}}, \ and\ \bibinfo {author} {\bibfnamefont {M.~J.}\ \bibnamefont {Alava}},\ }\href {\doibase 10.1088/1361-651X/ac9d54} {\bibfield  {journal} {\bibinfo  {journal} {Modelling and Simulation in Materials Science and Engineering}\ }\textbf {\bibinfo {volume} {30}},\ \bibinfo {pages} {085010} (\bibinfo {year} {2022})}\BibitemShut {NoStop}%
\bibitem [{\citenamefont {Bacon}\ \emph {et~al.}(2009)\citenamefont {Bacon}, \citenamefont {Osetsky},\ and\ \citenamefont {Rodney}}]{bacon2009dislocation}%
  \BibitemOpen
  \bibfield  {author} {\bibinfo {author} {\bibfnamefont {D.}~\bibnamefont {Bacon}}, \bibinfo {author} {\bibfnamefont {Y.~N.}\ \bibnamefont {Osetsky}}, \ and\ \bibinfo {author} {\bibfnamefont {D.}~\bibnamefont {Rodney}},\ }\href@noop {} {\bibfield  {journal} {\bibinfo  {journal} {Dislocations in solids}\ }\textbf {\bibinfo {volume} {15}},\ \bibinfo {pages} {1} (\bibinfo {year} {2009})}\BibitemShut {NoStop}%
\end{thebibliography}%
\bibliographystyle{apsrev4-1}

\end{document}